\documentclass[fleqn,usenatbib]{mnras}

\usepackage{newtxtext,newtxmath}
\usepackage[T1]{fontenc}

\DeclareRobustCommand{\VAN}[3]{#2}
\let\VANthebibliography\thebibliography
\def\thebibliography{\DeclareRobustCommand{\VAN}[3]{##3}\VANthebibliography}

\usepackage{graphicx}	% Including figure files
\usepackage{amsmath}	% Advanced maths commands

\usepackage{amssymb}	% Extra maths symbols
\usepackage{subfigure}
\usepackage[normalem]{ulem}
\usepackage{tikz,hyperref}
\usepackage[switch]{lineno}
\newcommand{\teff}{\mbox{$T_{\rm eff}$}} \newcommand{\logg}{{\rm{log}~$g$}}
\newcommand{\feh}{{\rm [Fe/H]}} 
\newcommand{\ebv}{$E(B-V)$}

\usepackage{threeparttable}

\DeclareRobustCommand{\orcidicon}{%
	\begin{tikzpicture}
	\draw[lime, fill=lime] (0,0) 
	circle [radius=0.16] 
	node[white] {{\fontfamily{qag}\selectfont \tiny ID}};
	\draw[white, fill=white] (-0.0625,0.095) 
	circle [radius=0.007];
	\end{tikzpicture}
	\hspace{-2mm}
}

\foreach \x in {A, ..., Z}{\expandafter\xdef\csname orcid\x\endcsname{\noexpand\href{https://orcid.org/\csname orcidauthor\x\endcsname}
			{\noexpand\orcidicon}}
}

\title{The miniJPAS survey: stellar atmospheric parameters from 56 optical filters}

\author[Yuan et al]{H.-B. Yuan,$^{1,2}$\thanks{E-mail: yuanhb@bnu.edu.cn}\orcidA{}
L. Yang,$^{3}$\orcidB{}
P. Cruz,$^{4,5}$
F. Jim\'enez-Esteban,$^{4,5}$
S. Daflon,$^{6}$
V.~M.~Placco,$^{7}$
S.~Akras,$^{8}$\orcidG{}\newauthor
E.~J.~Alfaro,$^{9}$
C.~Andr\'es Galarza,$^{6}$
D.~R.~Gon\c calves,$^{10}$
F.-Q. Duan,$^{3}$
J.-F. Liu,$^{11,12}$
J.~Laur,$^{13}$
E.~Solano,$^{4,5}$\newauthor
M.~Borges Fernandes,$^{6}$
A.~J.~Cenarro,$^{14}$
A.~Mar\'{\i}n-Franch,$^{14}$
J.~Varela,$^{14}$
A.~Ederoclite,$^{14}$
Carlos L\'opez-Sanjuan,$^{14}$\newauthor
R.~Abramo,$^{15}$
J.~Alcaniz,$^{6}$
N.~Ben{\'i}tez,$^{9}$
S.~Bonoli,$^{14,16,17}$
D.~Crist\'obal-Hornillos,$^{14}$
R.~A.~Dupke,$^{6,18,19}$\newauthor
Antonio Hern\'an-Caballero,$^{14}$
C.~Mendes de Oliveira,$^{20}$
M.~Moles,$^{14}$
L.~Sodr\'e Jr.,$^{20}$
H\'ector V\'azquez Rami\'o,$^{14}$\newauthor
K.~Taylor,$^{21}$
\\
$^{1}$Institute for Frontiers in Astronomy and Astrophysics, Beijing Normal University,  Beijing 102206, China\\
$^{2}$Department of Astronomy, Beijing Normal University No.19, Xinjiekouwai St, Haidian District, Beijing, 100875, China\\
$^{3}$College of Artificial Intelligence, Beijing Normal University No.19, Xinjiekouwai St, Haidian District, Beijing, 100875, China\\
$^{4}$Depto. de Astrof{\'i}sica, Centro de Astrobiolog{\'i}a (INTA-CSIC), ESAC campus, Camino Bajo del Castillo s/n, E-28692, Villanueva de la Ca{\~n}ada, Spain\\
$^{5}$Spanish Virtual Observatory (SVO), E-28692, Villanueva de la Ca{\~n}ada, Spain\\
$^{6}$Observat\'orio Nacional - MCTI (ON), Rua Gal. Jos\'e Cristino 77, S\~ao Crist\'ov\~ao, 20921-400 Rio de Janeiro, Brazil\\
$^{7}$NSF’s NOIRLab, 950 N. Cherry Ave., Tucson, AZ 85719, USA\\
$^{8}$ Institute for Astronomy, Astrophysics, Space Applications and Remote Sensing, National Observatory of Athens, GR 15236 Penteli, Greece\\
$^{9}$Instituto de Astrof\'{\i}sica de Andaluc\'{\i}a, IAA-CSIC, Glorieta de la Astronom\'{\i}a s/n, 18008 Granada, Spain\\
$^{10}$Observat\'orio do Valongo, Universidade Federal do Rio de Janeiro, Ladeira Pedro Antonio 43, Rio de Janeiro 20080-090, Brazil\\
$^{11}$National Astronomical Observatories, Chinese Academy of Sciences, 20A Datun Road, Chaoyang District, Beijing, China\\
$^{12}$University of Chinese Academy of Sciences, Yuquan Road, Shijingshan District, Beijing, China\\
$^{13}$Tartu Observatory, University of Tartu, Observatooriumi 1, 61602 T\~oravere, Estonia\\
$^{14}$Centro de Estudios de F\'{\i}sica del Cosmos de Arag\'on (CEFCA), Unidad Asociada al CSIC, Plaza San Juan 1, 44001 Teruel, Spain\\
$^{15}$Instituto de F\'{\i}sica, Universidade de S\~ao Paulo, Rua do Mat\~ao 1371, CEP 05508-090, S\~ao Paulo, Brazil\\
$^{16}$Donostia International Physics Centre (DIPC), Paseo Manuel de Lardizabal 4, 20018 Donostia-San Sebastián, Spain\\
$^{17}$IKERBASQUE, Basque Foundation for Science, 48013, Bilbao, Spain\\
$^{18}$University of Michigan, Department of Astronomy, 1085 South University Ave., Ann Arbor, MI 48109, USA\\
$^{19}$University of Alabama, Department of Physics and Astronomy, Gallalee Hall, Tuscaloosa, AL 35401, USA\\
$^{20}$Instituto de Astronomia, Geof\'{\i}sica e Ci\^encias Atmosf\'ericas, Universidade de S\~ao Paulo, 05508-090 S\~ao Paulo, Brazil\\
$^{21}$Instruments4, 4121 Pembury Place, La Canada Flintridge, CA, 91011, USA
}

\date{Accepted 2022. Received 2022; in original form ZZZ}

\pubyear{2022}

\begin{document}
%\linenumbers
\label{firstpage}
\pagerange{\pageref{firstpage}--\pageref{lastpage}}
\maketitle

\begin{abstract}
With a unique set of 54 overlapping narrow-band and 
two broader filters covering the entire optical range,
the incoming {\it Javalambre-Physics of the Accelerating Universe Astrophysical Survey} (J-PAS) will provide a great opportunity for stellar physics and near-field cosmology.
In this work, we use the miniJPAS data in 56 J-PAS filters and 4 complementary SDSS-like filters to explore and prove the potential of the J-PAS filter system in characterizing stars and deriving their atmospheric parameters. 
We obtain estimates for the effective temperature with a good precision ($<$150\,K) from spectral energy distribution fitting.
We have constructed the metallicity-dependent stellar loci in 59 colours for the miniJPAS FGK dwarf stars, after correcting certain systematic errors in flat-fielding. The very blue colours, including $uJAVA - r$, $J0378-r$, $J0390-r$, $uJPAS-r$, show the strongest metallicity dependence, around 0.25\,mag/dex.  The sensitivities decrease to about 0.1 mag/dex for the $J0400-r$, $J0410-r$, and $J0420-r$ colours. The locus fitting residuals show peaks at the $J0390, J0430, J0510$, and $J0520$ filters, suggesting that individual elemental abundances such as [Ca/Fe], [C/Fe], and [Mg/Fe] can also be determined from the J-PAS photometry. Via stellar loci, we have achieved a typical metallicity precision of 0.1\,dex. The miniJPAS filters also demonstrate strong potential in discriminating dwarfs and giants, particularly the $J0520$ and $J0510$ filters.
Our results demonstrate the power of the J-PAS filter system in stellar parameter determinations and the huge potential of the coming J-PAS survey in stellar and Galactic studies.
\end{abstract}


%%%%%%%%%%%%%%%%%%%%%%%%%%%%%%%%%%%%%%%%%%%%%%%%%%

%%%%%%%%%%%%%%%%% BODY OF PAPER %%%%%%%%%%%%%%%%%%

\section{Introduction}

The {\it Javalambre-Physics of the Accelerating Universe Astrophysical Survey} \citep[J-PAS;][]{Benitez2014} aims to image 8500 square degrees of the northern sky with a unique set of 54 overlapping narrow-band filters and 2 broad-band filters, with a dedicated 2.5-m telescope, JST/T250, at the {\it Observatorio Astrofísico de Javalambre} \citep[OAJ;][]{Cenarro2014}. The  narrow-band filters have a tpyical FWHM of 145\,\AA, covering the entire optical range from 3785 to 9100\,\AA. 
Driven by accurate redshift estimates of galaxies over a wide range of epochs, the J-PAS survey will effectively deliver a low-resolution spectra for every single pixel of the 8500 deg$^2$ sky area observed, providing valuable datasets for not only cosmology and galaxy evolution, but also stellar physics and near-field cosmology. 

As a pathfinder for J-PAS, the miniJPAS survey \citep{Bonoli2021} has imaged the famous Extended Groth Strip (EGS) field of about 1\,deg$^2$, with the 
JPAS-Pathfinder (JPF) camera. 
In addition to test the performance of the telescope, the miniJPAS data offers a great opportunity to prove the potential of the J-PAS filter system. % some examples here. 
In this work, we explore the power of these filters in characterizing stars and deriving their fundamental parameters via different techniques.

Metallicity-dependent stellar locus \citep{Yuan2015a} provides a simple yet non-trivial tool to investigate the dependencies of colours on metallicity and to determine the metallicity for millions of stars from multi-band photometry. \citet{Yuan2015b} applied this technique to the re-calibrated Sloan Digital Sky Survey \citep[SDSS;][]{York2000} SDSS/Stripe 82 \citep{Yuan2015c} and obtained metallicities for half a million FGK dwarf stars, with a typical error of $\sigma$[Fe/H] $\sim$ 0.1-0.2\,dex. Using the same data, \citet{Zhang2021} obtained metallicity-dependent stellar loci for red-giant stars and used them to derive metallicities of giants to a precision of $\sigma$[Fe/H] $\sim$ 0.20-0.25\,dex, and to discriminate metal-poor red giants from main-sequence stars based on SDSS photometry. With the recalibrated SkyMapper Southern Survey (SMSS) DR2 \citep{Huang2021}, \citet{Huang2022} determined metallicities for over 24 million stars with a similar technique, achieving a precision comparable to or slightly better than that derived from spectroscopy for stars with metallicities as low as [Fe/H] $\sim -3.5$. 
Based on corrected broad-band Gaia Early Data Release 3 \citep[Gaia EDR3;][]{Gaia2016,Gaia2021} colours alone by \citet{Niu2021b} and \citet{2021ApJ...908L..24Y}, and a careful reddening correction, \citet{Xu2022} determined reliable metallicity estimates for a magnitude-limited sample of about 27 million stars down to [Fe/H] = $-2.5$. 

In this work, we first determine effective temperatures via spectral energy distribution (SED) fitting, based on miniJPAS photometry. We then investigate the dependencies of miniJPAS colours on the metallicity and perform \feh\ estimates for FGK dwarf stars via the metallicity-dependent stellar loci. At last, we explore giant/dwarf classifications using a machine learning technique to identify the most sensitive colours. 
Note that measurements of surface gravity are not discussed in this work.

This paper is organized as follows: Section\,\ref{data} describes the miniJPAS data used in this work. Section\,\ref{vosa} presents the stellar characterization using the available photometric data. The metallicity-dependent stellar loci are determined and discussed in Section \,\ref{sl}. Metallicity estimates are described and compared in Section\,\ref{feh} and Section \,\ref{giant} discusses giant/dwarf classifications. Finally, we summarize our work in Section \,\ref{summary}.

%--------------------------------------------------------------------

\section{MiniJPAS Data} \label{data}

%\subsection{miniJPAS}

The miniJPAS was carried out at OAJ with the JPF camera mounted on the JST/T250 telescope. JPF was the first scientific instrument of the JST/T250, before the arrival of the Javalambre Panoramic Camera \citep[JPCam;][]{Taylor2014,Mar2017}. The JPF has a single 9200$\times$9200 CCD, located at the center of the JST250 field of view (FoV), with a pixel scale of 0.23\,arcsec\,pixel$^{-1}$, providing an effective FoV of 0.27\,deg$^2$.

In addition to the 54 narrow-band filters, ranging from 3785 to 9100\,\AA, two broader filters, $J0348$ (also known as $uJAVA$, that covers the ultraviolet edge) and $J1007$ (that covers the red wings), and four SDSS-like broad-band filters ($uJPAS$, $g$, $r$, and $i$) are also used. The J-PAS filters have been optimized to deliver a low-resolution ($R \sim 60$) photo-spectrum of each imaged pixel/target, therefore probing a large number of key stellar absorption features. The technical description and characterization of the J-PAS filters can be found in \citet{Mar2012}. Detailed information of the filters as well as their transmission curves are available at the Spanish Virtual Observatory Filter Profile Service\footnote{\url{http://svo2.cab.inta-csic.es/theory/fps/index.php?mode=browse&gname=OAJ&gname2=JPAS}}.

The miniJPAS covers the Extended Groth Strip area at ($\alpha$, $\delta$) = (215, $+$53)\,deg with four tiles, with a total area of about 1 deg$^2$ \citep[see][for further details]{Bonoli2021}. All data collected are processed and calibrated by the Data Processing and Archiving Unit group at CEFCA\footnote{Centro de Estudios de F{\'i}sica del Cosmos de Arag{\'o}n (Spain).}. The image depths achieved -- 5$\sigma$ in a 3\,arcsec circular aperture -- are deeper than 22\,mag for filters bluewards of 7500\AA\ and about 22\,mag for longer wavelengths. The images and catalogues are publicly available on the J-PAS website\footnote{\url{http://www.j-pas.org/datareleases/minijpas_public_data_release_pdr201912}}. 
The 6\,arcsec-aperture magnitudes in the ``dual mode'' catalogues are used in this work, along with aperture corrections. In the ``dual mode'', the $r$-band images were used to detect targets with SExtractor \citep{Bertin1996}, and the apertures defined then were also used in other filters.

%\subsection{Sample selection} 

We select stars from the miniJPAS database that fulfill the following criteria: $r <$ 22\,mag and CLASS\_STAR\footnote{Morphological classification parameter provided by
SExtractor} $>$ 0.6. A number of 2895 stars are selected, including 1566 and 749 stars brighter than 20 and 18 magnitude in the $r$ band, respectively. The selected stars are cross-matched with the LAMOST DR7 \citep{Luo2015} with a match radius of 1 arcsec, 161 common stars are found, including 10 giant stars. The selected stars are also cross-matched with the SDSS/SEGUE catalog \citep{Rockosi2022}, with only 31 stars found. 
We also cross-match the selected stars with the J-PLUS DR1 stellar parameters catalogue from \citet{Yang2022}, 682 stars of $r < 18$ are found.
Given the small number of common stars between the miniJPAS catalogue with LAMOST or SDSS spectroscopic catalogues, we have adopted different approaches to characterize the stellar miniJPAS sample, which will be described in the following sections.

\section{Stellar effective temperature from SEDs} 
\label{vosa}

\subsection{SED fitting with VOSA}

\begin{figure}
	\centering
	\includegraphics[width=0.95\columnwidth]{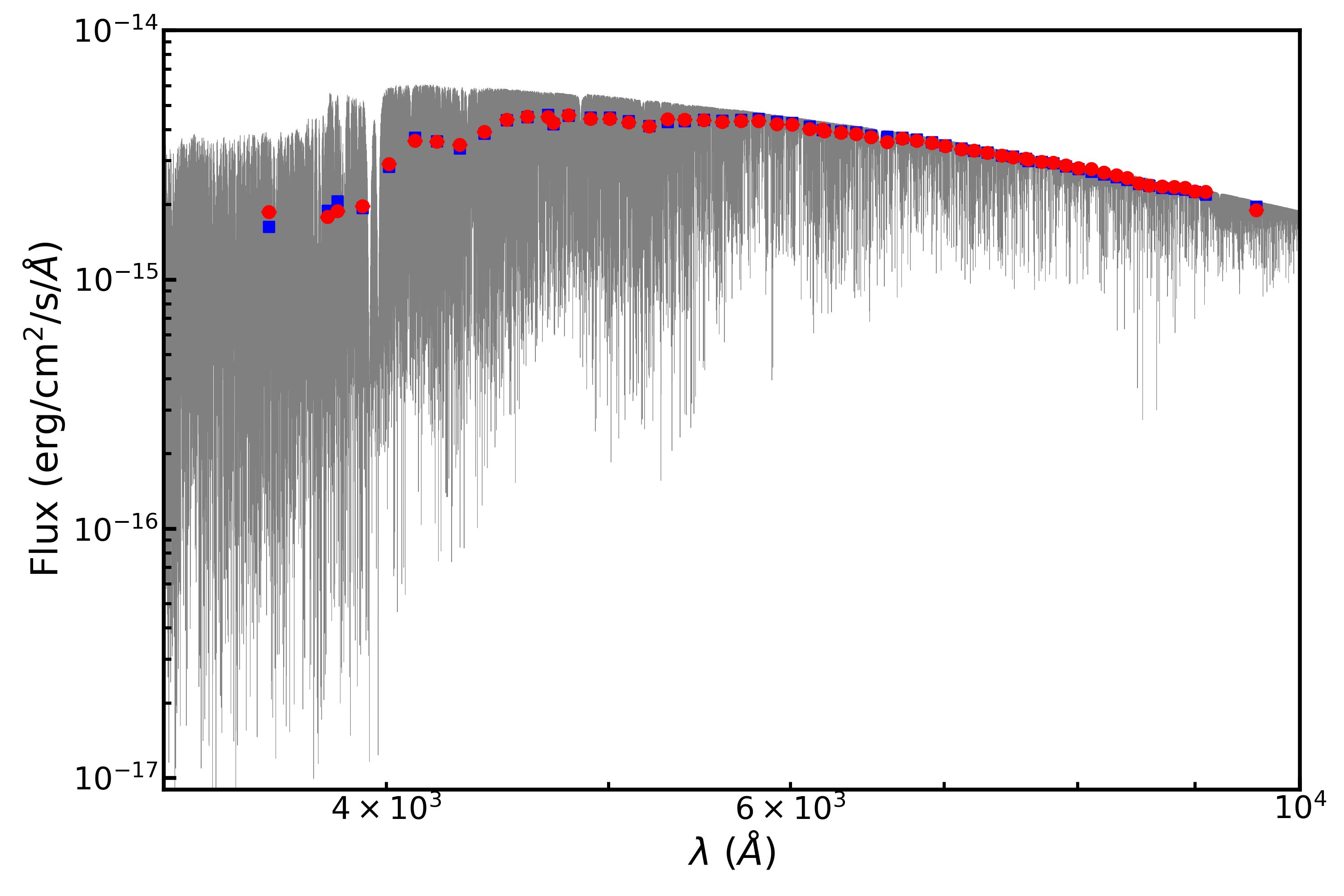}
	\includegraphics[width=0.95\columnwidth]{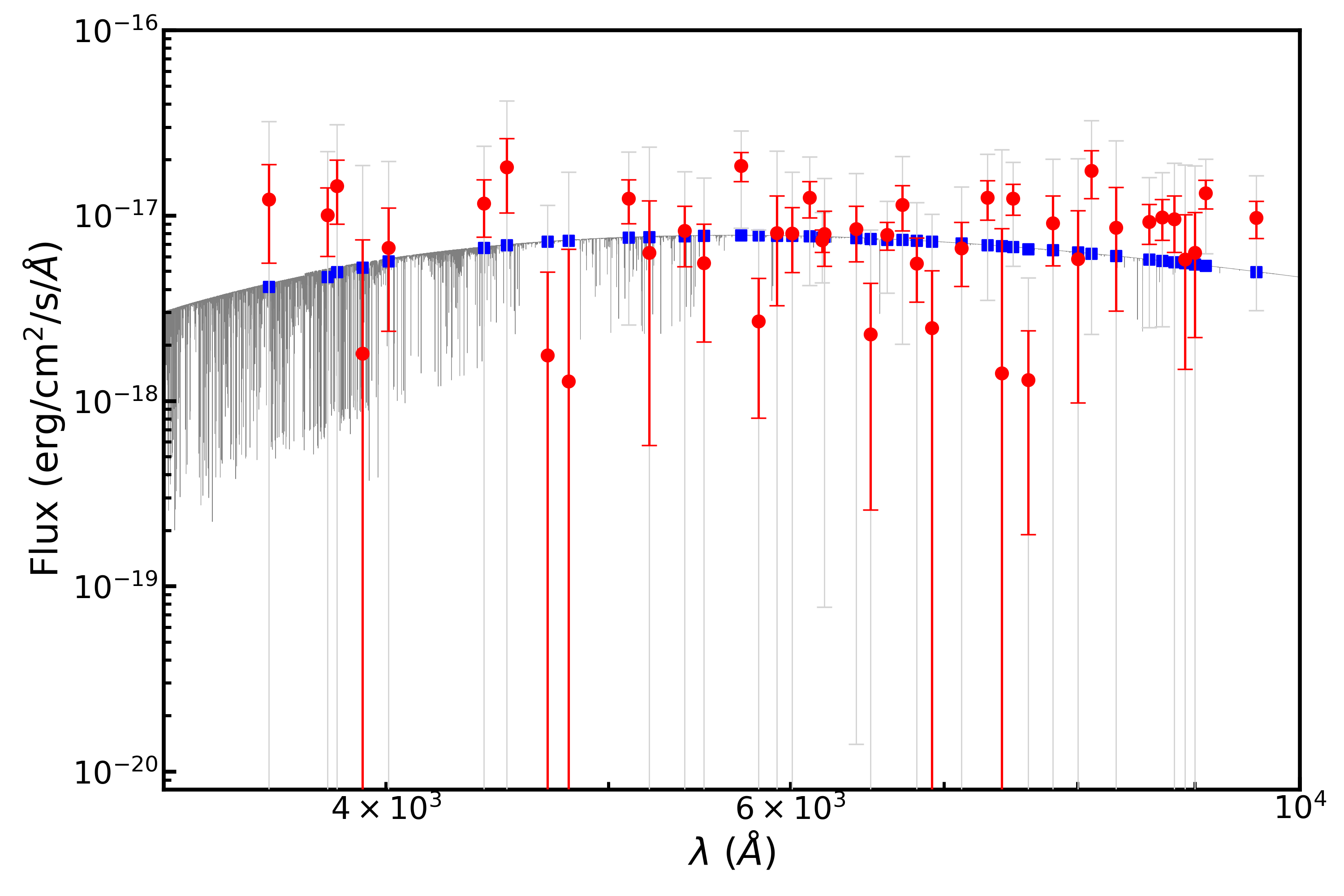}
	\caption{SED fittings obtained with VOSA. The observed J-PAS photometric data are shown in red. The synthetic photometry obtained from the best-fitting BT-Settl model is presented as blue squares, and the corresponding spectra represented by a solid grey line. The light-grey vertical bars are the 3-$\sigma$ uncertainties. {\sl Top panel:} SED best-fitting model for the 2406-6604 object, with $Vgf_b$ of $\sim$0.07. {\sl Bottom panel:} SED best-fitting model for the 2243-8717 object as an example of a rejected source, with $Vgf_b$ of $\sim$3.02, just above the imposed threshold (see text).} %Note that these plots were generated by VOSA.}
	\label{fig:SEDexample}
\end{figure}

\begin{figure}
	\centering
	\includegraphics[width=0.97\columnwidth]{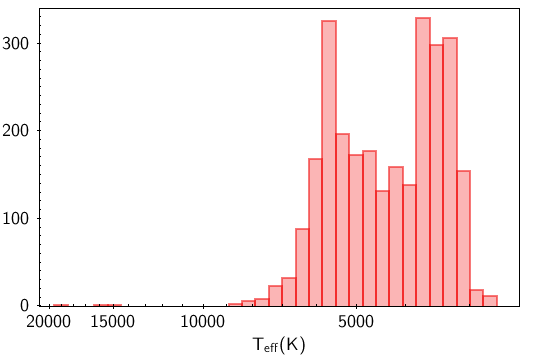}%	
	\caption{Histograms showing the distribution of $T_{\rm eff}$ %(middle panel) and $\log g$ (bottom panel) 
	obtained from SED fitting with VOSA for the sample with a good fit (Vgf$_b<3$). 
	}
	\label{fig:histVOSA}
\end{figure}

We used the Virtual Observatory Sed Analyzer\footnote{\url{http://svo2.cab.inta-csic.es/theory/vosa/}} \citep[VOSA;][]{Bayo2008} and the SEDs constructed from the miniJPAS photometry data, to obtain the stellar 
effective temperatures (\teff)
of the objects in the miniJPAS survey via atmospheric model fitting. We used the BT-Settl spectra models by \citet[][and references therein]{Allard2011}, computed with a cloud model, for all available temperatures -- from 400\,K to 70\,000\,K, with a pace of 100\,K till the 7\,000\,K, then a pace of 200\,K till the 12\,000\,K, increasing the pace to 500\,K till the 12\,000\k, and with a final pace of 1000\,K for temperatures larger than 20\,000\,K -- with \logg\ varying from 2.0 to 6.0\,dex and the metallicity ([M/H]) from $-$4.0 to $+$0.5, both with a pace of 0.5. VOSA also allows to work with a range of extinctions. We set this range from $A_v$=0 to 0.1, the maximum Milky Way extinction in the region of the sky covered by the miniJPAS \citep{Bonoli2021}. VOSA unreddens the miniJPAS SEDs using 20 values of $A_v$ within this range in steps of 0.005\,mag, and selected the $A_v$ value which unreddened SED best fit the model. 
It is worth mentioning that fit parameter uncertainties are estimated in VOSA by performing a Monte Carlo simulation with a 100 iterations. Taking the observed SED as the starting point, VOSA generates 100 virtual SEDs introducing a Gaussian random noise for each point proportional to the observational error. VOSA obtains the best fit for the 100 virtual SEDs with noise and makes the statistics for the distribution of the obtained values for each parameter. The standard deviation for this distribution will be reported as the uncertainty for the parameter if its value is larger that half the grid step for this parameter.

The complete miniJPAS stellar sample contains 2\,895 objects. We constructed the SEDs only with good photometric data (individual filter photometry flag$<$4). This translated in 2\,871 miniJPAS SEDs constructed using between 1 and 60 photometry bands (including the four SDSS-like filters). However, considering the variables on the fitting procedure, a minimum of 6 photometric points are needed to perform the fit, which reduced the sample to 2\,858 SEDs, most of them (>94\%) with at least 50 photometric points.

From the 2\,858 analysed SEDs, we found a good-fit solution -- defined by a modified reduced $\chi^2$ value ($Vgf_b$\footnote{Vgfb: Modified reduced $\chi^{2}$, calculated by forcing $\sigma(F_{obs})$ to be larger than $0.1\times F_{obs}$, where $\sigma(F_{obs})$ is the error in the observed flux ($F_{obs}$). This can be useful if the photometric errors of any of the catalogues used to build the SED are underestimated. Vgfb smaller than 10--15 is often perceived as a good fit. \url{http://svo2.cab.inta-csic.es/theory/vosa/help/star/fit/}.}) of less than 3 -- for 2\,732 of them, representing $\sim$95\% of the miniJPAS analysed sample. 
To illustrate, Figure \ref{fig:SEDexample} shows two examples of obtained SED fittings, the one with the lowest $Vgf_b$, and a rejected solution, close to the cutoff limit of $Vgf_b$=3.

The temperatures obtained range from 2\,700 to 19\,000\,K. Figure \ref{fig:histVOSA} shows the \teff\ 
distribution obtained from VOSA for the sample with best SED fitting results -- %2\,717 objects with $Vgf_b<3$.
2\,732 objects with $Vgf_b<3$. 
The distribution is consistent with what is expected from a field stellar sample, in which dwarfs and giants are present \citep{VanderSwaelmen2022}.

\subsection{The LAMOST control sample}

In order to evaluate both the accuracy of the fitting method and the reliability of the derived effective temperature, we compare the VOSA results with the effective temperature derived from spectroscopy. 
We cross-matched the miniJPAS stellar sample with the LAMOST DR7 
catalogue \citep{Luo2015}, finding 151 LAMOST objects for which we obtained a good fit with VOSA. 

\begin{figure}
	\centering
\includegraphics[width=0.95\columnwidth]{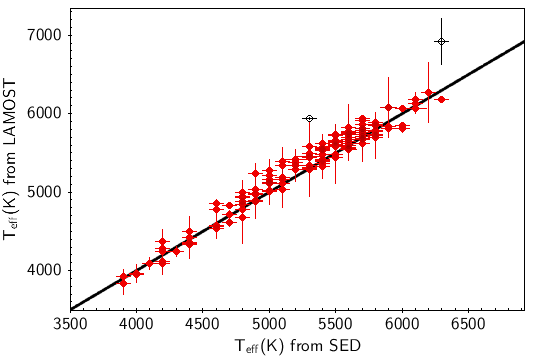}
\includegraphics[width=0.96\columnwidth]{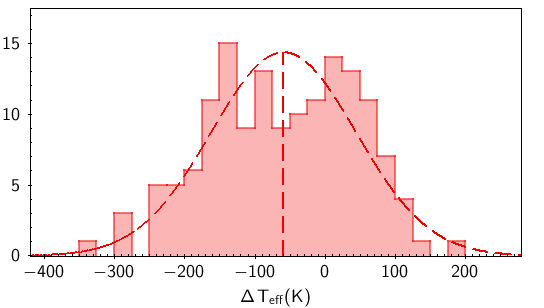}
	\caption{Comparison between effective temperatures obtained from LAMOST spectroscopic data and from miniJPAS SED fitting with VOSA. The black-empty circles are the 3-$\sigma$ clipped outliers. {\sl Top panel:} Temperatures from LAMOST DR7 (y-axis) and those from VOSA (x-axis), where the solid line is the identity function. {\sl Bottom panel:} Distribution of \teff\ differences between VOSA and LAMOST. Dashed line corresponds to a Gaussian fit of the distribution, with a mean \teff\ value of $-$60\,K and a $\sigma\sim$103\,K, approximately.} 
	\label{fig:TeffVOSA}
\end{figure}

Figure \ref{fig:TeffVOSA} (top panel) shows the comparison between the $T_{\rm eff}$ obtained by miniJPAS SED fitting with VOSA, and those obtained from spectroscopy by LAMOST. We found a good agreement, within uncertainties, with the SED fitting temperatures slightly cooler than the spectroscopic ones.

Figure \ref{fig:TeffVOSA} (bottom panel) shows the distribution of the difference in $T_{\rm eff}$ between the values obtained from the miniJPAS SED fitting with VOSA and those from LAMOST spectroscopy, after removing the 3-$\sigma$ clipped outliers. We fitted the distribution to a Gaussian curve, resulting in a mean value of $-$60\,K and a standard deviation of 103\,K, approximately. From this, and considering that the LAMOST $T_{\rm eff}$ uncertainties have a median value of $\sim$60\,K, we can conservatively estimate a typical $T_{\rm eff}$ uncertainty lower than 150\,K in the VOSA miniJPAS SED fitting, in the range of temperatures in common. This result validates the miniJPAS data set and the SED fitting methodology with VOSA as a reliable way to obtain stellar effective temperatures for thousands of sources, without the need of more resource demanding spectroscopic observations. 

From the analysed LAMOST sample, we reached the conclusion that even with the amazing collection of 56 narrow-band filters, there is not enough resolution in the SED to derive surface gravities and metallicities with good precision. This result was expected as SEDs have much lower sensitivity to changes in logg and metallicity. Therefore, other approaches are presented below.

\subsection{J-PLUS stellar sample}

\begin{figure}
	\centering
    \includegraphics[width=0.95\columnwidth]{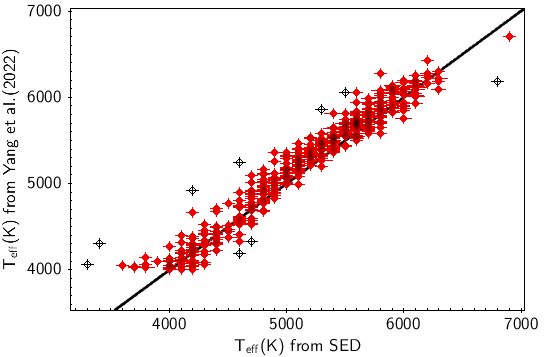}
    \includegraphics[width=0.96\columnwidth]{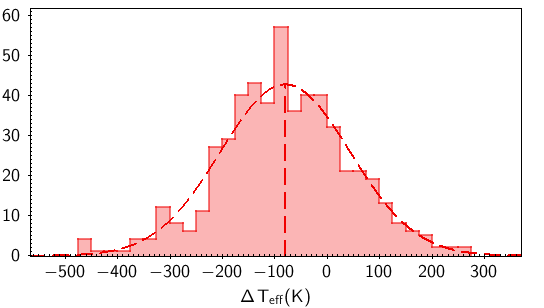}
    \caption{Comparison between effective temperatures from VOSA (this work) to those from \citet{Yang2022} for the common subset, with 542 stars. The black-empty circles are the 3-$\sigma$ clipped outliers. {\sl Top panel:} Temperatures from previous work \citep{Yang2022} (y-axis) and those from VOSA (x-axis), where the solid line is the identity function. {\sl Bottom panel:} Distribution of \teff\ differences between VOSA and \citet{Yang2022}. Dashed line corresponds to a Gaussian fit of the distribution, with a mean \teff\ value of $-$81\,K and a $\sigma\sim$124\,K, approximately.} 
   
    \label{fig:TeffYang}
\end{figure}

\begin{figure}
	\centering
\includegraphics[width=\columnwidth]{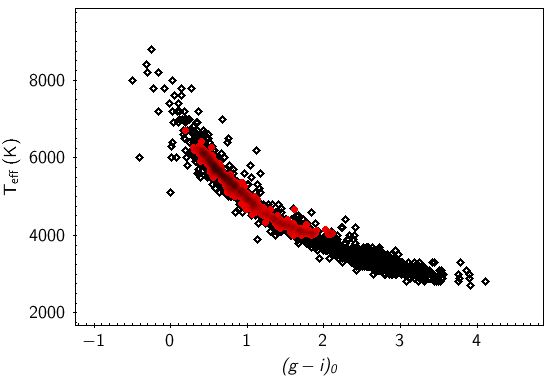}
    \caption{$T_{\rm eff}$ versus {\it (g$-$i)$_0$} diagram. The black diamonds show the temperatures obtained from VOSA analysis (this work) for 2\,503 objects -- only those objects that presented good g- and i-band photometry (flag\,$<$\,4; ERR\_APER6\,$<$\,1) and VOSA visual good fit (Vgf$_b<3$) are shown (see text for details). The red circles represent the temperatures obtained by \citet{Yang2022} for the objects in common (542 stars). Only objects remaining after the 3-$\sigma$ clipping are shown.}
    \label{fig:Teffcolour}
\end{figure}

Given the small numbers of stars in common between the miniJPAS catalogue and the LAMOST/SDSS spectroscopic catalogues, we further used the J-PLUS stellar parameters from \citet{Yang2022} to compare with our miniJPAS SED fitting results.
The J-PLUS DR1 catalogue was re-calibrated with the stellar colour regression method \citep{Yuan2015c,Huang2021,Niu2021a,Niu2021b,HuangYuan2022,XiaoYuan2022}, achieving a calibration precision of 2-5\,mmag in all 12 J-PLUS filters. 
Based on the re-calibrated JPLUS DR1 and {\it Gaia} DR2 \citep{Gaia2018}, \citet{Yang2022} trained 8 cost-sensitive neural networks, the CSNet, to learn the non-linear mapping from stellar colours to their labels independently. 
The stellar parameter achieved internal precisions were of 55\,K for \teff, about 0.15\,dex for \logg, and 0.07\,dex for \feh, respectively. Abundance for [$\alpha$/Fe] and 4 individual elements ([C/Fe], [N/Fe], [Mg/Fe], and [Ca/Fe]) were also estimated, with a precision of 0.04-0.08\,dex. The catalogue containing about 2 million stars is publicly available\footnote{\url{http://www.j-plus.es/ancillarydata/index}}.

We cross-matched J-PLUS DR1 stellar parameters catalogue from \citet{Yang2022} with the miniJPAS sample with good-fitting result from VOSA (2\,732 objetcs). A total of 542 stars were found in common. The obtained VOSA temperatures (this work) are in a very good agreement with the previous results by \citet{Yang2022}, as illustrated in Figure \ref{fig:TeffYang}, where both results are compared. The $T_{\rm eff}$ obtained from the miniJPAS SED fitting with VOSA are presented in the abscissa axis. The temperatures by \citet{Yang2022} are presented in the ordinate axis (top panel), showing the accordance in respect to the identity function (black line). The difference between both results, this work and the results by \citet{Yang2022}, are also shown (bottom panel). After applying a 3-$\sigma$ clipping, we found a mean difference of $-$81\,K, with a standard deviation ($\sigma$) of 124\,K, approximately. 
These results are in complete agreement, within the estimated errors, to those obtained in the comparison to LAMOST data, being VOSA \teff\ slightly cooler in the 4800$-$5800\,K range. Probably, this trend is due to different systematics between the VOSA and J-PLUS/LAMOST results,
%to some systematics in J-PLUS results, 
since \citet{Yang2022} used LAMOST data for training the artificial neural networks and the same behavior is seen when comparing the SED \teff\ obtained from VOSA to those from LAMOST spectra (see fig. \ref{fig:TeffVOSA}, top panel).

The \teff\ versus {\it (g$-$i)$_0$} diagram is presented in Figure \ref{fig:Teffcolour}, where black diamonds show the temperatures obtained from VOSA analysis for the miniJPAS SEDs. Note that only those stars with \teff\ $<$ 10\,000\,K are presented. For comparison, the temperatures obtained by \citet{Yang2022} are also shown, represented by red filled circles. This figure shows that VOSA \teff\ follows a similar trend with the {\it (g$-$i)$_0$} colour to the one found by \citet{Yang2022}, but extending to a wider range of temperatures. This, again, validate our methodology to derive effective temperatures for thousands of sources from miniJPAS SEDs with VOSA.

\section{Metallicity-dependent stellar loci of FGK dwarf stars} \label{sl}

We also used the \citet{Yang2022} J-PLUS DR1 stellar parameter estimations to determine the metallicity-dependent stellar loci of the miniJPAS colours for 
FGK dwarf stars. After requiring $r < 18, 0.3 < (g-i)_0$ < 1.6, 4000 < \teff~< 7000 K, $\log g$ > 4.0, and flag = 0 in all the miniJPAS filters, 310 FGK dwarf stars of high-quality photometry and 
stellar parameters are selected. These stars, including 10 with [Fe/H] $< -1.5$, are used to determine the metallicity-dependent stellar loci of the miniJPAS colours.

\begin{table*}
    \centering
    \caption{Fitting coefficients of flat-fielding errors in 56 J-PAS filters + 4 SDSS-like filters}
    %different filters.}
    \begin{threeparttable}
    \begin{tabular}{lrrrrrr}
    \hline
    Filter & $a_0$ &  $a_1$ &  $a_2$ &  $a_3$ & $a_4$ & $a_5$\\
    \hline 	
    uJAVA  &    $-$0.0372  &    1.65e-06  &    $-$4.47e-10  &    1.20e-05  &    1.29e-10  &    $-$6.57e-10  \\
    J0378  &    $-$0.0778  &    8.89e-06  &    $-$4.46e-10  &    1.95e-05  &    $-$4.22e-10  &    $-$1.22e-09  \\
    J0390  &    $-$0.0770  &    1.77e-06  &    2.38e-10  &    2.01e-05  &    $-$1.05e-10  &    $-$9.81e-10  \\
    J0400  &    $-$0.0391  &    2.75e-06  &    $-$2.59e-10  &    9.07e-06  &    2.19e-10  &    $-$4.74e-10  \\
    J0410  &    $-$0.0256  &    $-$1.84e-06  &    2.72e-10  &    5.84e-06  &    4.05e-10  &    $-$2.76e-10  \\
    J0420  &    $-$0.0286  &    4.19e-07  &    4.51e-11  &    7.84e-06  &    2.53e-10  &    $-$5.75e-10  \\
    J0430  &    $-$0.0313  &    4.15e-06  &    $-$2.85e-10  &    4.71e-06  &    1.04e-10  &    $-$1.16e-10  \\
    J0440  &    $-$0.0291  &    2.92e-06  &    $-$1.16e-10  &    5.79e-06  &    $-$5.99e-11  &    $-$2.30e-10  \\
    J0450  &    $-$0.0093  &    $-$2.30e-06  &    2.15e-10  &    2.19e-06  &    2.81e-10  &    $-$3.94e-11  \\
    J0460  &    $-$0.0137  &    7.36e-07  &    $-$8.81e-11  &    3.14e-06  &    9.87e-11  &    $-$1.36e-10  \\
    J0470  &    $-$0.0173  &    3.78e-06  &    $-$4.01e-10  &    3.67e-06  &    4.52e-11  &    $-$2.91e-10  \\
    J0480  &    $-$0.0214  &    4.08e-06  &    $-$3.86e-10  &    2.27e-06  &    3.70e-11  &    6.99e-11  \\
    J0490  &    $-$0.0044  &    6.19e-07  &    $-$4.80e-11  &    1.12e-06  &    $-$6.25e-11  &    $-$4.02e-11  \\
    J0500  &    $-$0.0078  &    1.74e-07  &    6.68e-11  &    1.30e-06  &    $-$2.29e-10  &    1.60e-10  \\
    J0510  &    $-$0.0214  &    4.00e-06  &    $-$2.89e-10  &    3.74e-06  &    $-$2.37e-10  &    $-$6.02e-11  \\
    J0520  &    $-$0.0166  &    $-$3.21e-07  &    5.52e-11  &    4.82e-06  &    $-$5.34e-11  &    $-$1.59e-10  \\
    J0530  &    $-$0.0021  &    $-$6.36e-07  &    3.83e-11  &    8.46e-07  &    $-$4.93e-11  &    4.30e-11  \\
    J0540  &    $-$0.0185  &    4.97e-06  &    $-$4.84e-10  &    1.39e-06  &    1.46e-10  &    $-$4.10e-11  \\
    J0550  &    $-$0.0078  &    2.34e-06  &    $-$5.41e-11  &    $-$1.22e-06  &    $-$1.33e-10  &    2.84e-10  \\
    J0560  &    0.0020  &    3.53e-07  &    $-$1.38e-10  &    1.51e-06  &    $-$1.74e-10  &    $-$1.53e-10  \\
    J0570  &    $-$0.0001  &    1.53e-06  &    $-$1.64e-10  &    $-$2.17e-06  &    $-$2.55e-11  &    3.07e-10  \\
    J0580  &    $-$0.0158  &    4.04e-06  &    $-$2.89e-10  &    1.06e-06  &    $-$1.67e-10  &    1.42e-10  \\
    J0590  &    $-$0.0026  &    1.22e-06  &    1.20e-11  &    $-$1.41e-06  &    $-$1.12e-10  &    2.29e-10  \\
    J0600  &    0.0033  &    $-$7.71e-07  &    $-$2.43e-11  &    $-$5.27e-07  &    7.03e-11  &    6.61e-11  \\
    J0610  &    $-$0.0125  &    2.12e-06  &    $-$4.45e-11  &    2.15e-06  &    $-$1.14e-10  &    $-$1.29e-10  \\
    J0620  &    0.0043  &    8.08e-07  &    $-$9.63e-11  &    $-$3.05e-06  &    $-$4.09e-11  &    3.60e-10  \\
    J0630  &    $-$0.0030  &    1.24e-06  &    $-$5.87e-11  &    $-$1.70e-06  &    $-$1.39e-11  &    2.81e-10  \\
    J0640  &    $-$0.0017  &    1.12e-06  &    $-$1.87e-12  &    $-$1.95e-06  &    $-$8.76e-11  &    3.01e-10  \\
    J0650  &    $-$0.0058  &    3.17e-06  &    $-$2.46e-10  &    $-$1.27e-06  &    $-$1.04e-10  &    2.26e-10  \\
    J0660  &    0.0074  &    $-$6.19e-07  &    $-$4.25e-11  &    $-$2.64e-06  &    7.10e-13  &    3.29e-10  \\
    J0670  &    0.0035  &    5.15e-08  &    $-$7.59e-11  &    $-$1.53e-06  &    $-$2.65e-11  &    2.18e-10  \\
    J0680  &    0.0127  &    $-$5.86e-07  &    7.65e-11  &    $-$4.57e-06  &    $-$3.76e-11  &    3.78e-10  \\
    J0690  &    0.0103  &    $-$6.42e-07  &    9.98e-11  &    $-$4.43e-06  &    $-$2.26e-11  &    4.29e-10  \\
    J0700  &    0.0053  &    5.15e-07  &    $-$1.28e-10  &    $-$1.85e-06  &    $-$2.04e-10  &    3.11e-10  \\
    J0710  &    0.0005  &    2.02e-06  &    $-$1.49e-10  &    $-$2.33e-06  &    $-$1.30e-10  &    2.92e-10  \\
    J0720  &    0.0049  &    8.84e-08  &    $-$1.89e-11  &    $-$3.45e-06  &    $-$3.45e-11  &    4.66e-10  \\
    J0730  &    $-$0.0003  &    3.09e-06  &    $-$3.37e-10  &    $-$2.97e-06  &    7.73e-11  &    2.63e-10  \\
    J0740  &    0.0063  &    $-$8.02e-07  &    4.86e-12  &    $-$1.12e-06  &    8.82e-12  &    7.74e-11  \\
    J0750  &    0.0046  &    $-$1.65e-07  &    2.14e-11  &    $-$1.31e-06  &    1.47e-11  &    4.72e-11  \\
    J0760  &    0.0032  &    9.41e-07  &    $-$4.28e-11  &    $-$3.13e-06  &    4.20e-11  &    2.80e-10  \\
    J0770  &    0.0173  &    $-$1.51e-06  &    $-$3.35e-10  &    $-$2.64e-06  &    3.54e-12  &    3.49e-10  \\
    J0780  &    0.0108  &    2.76e-06  &    $-$2.33e-10  &    $-$6.05e-06  &    $-$3.44e-11  &    4.23e-10  \\
    J0790  &    $-$0.0045  &    1.65e-06  &    $-$2.32e-10  &    4.11e-07  &    6.03e-12  &    3.18e-11  \\
    J0800  &    0.0155  &    $-$7.40e-08  &    $-$2.09e-10  &    $-$5.53e-06  &    1.46e-10  &    4.69e-10  \\
    J0810  &    $-$0.0105  &    2.13e-06  &    $-$2.00e-10  &    4.77e-06  &    $-$2.58e-10  &    $-$4.24e-10  \\
    J0820  &    $-$0.0176  &    6.45e-06  &    $-$2.55e-10  &    2.52e-06  &    $-$3.56e-10  &    $-$3.64e-10  \\
    J0830  &    0.0028  &    4.10e-06  &    $-$1.38e-10  &    $-$6.03e-06  &    $-$3.26e-10  &    6.54e-10  \\
    J0840  &    0.0214  &    5.00e-07  &    $-$6.66e-11  &    $-$9.35e-06  &    $-$2.31e-10  &    9.89e-10  \\
    J0850  &    0.0140  &    4.48e-06  &    $-$7.33e-10  &    $-$8.61e-06  &    1.36e-10  &    7.96e-10  \\
    J0860  &    0.0003  &    $-$2.37e-07  &    1.36e-10  &    $-$2.41e-07  &    $-$1.17e-10  &    3.73e-11  \\
    J0870  &    0.0175  &    $-$1.52e-06  &    1.66e-10  &    $-$5.13e-06  &    7.05e-11  &    2.78e-10  \\
    J0880  &    0.0107  &    1.44e-06  &    $-$1.08e-10  &    $-$5.53e-06  &    3.56e-11  &    3.90e-10  \\
    J0890  &    $-$0.0007  &    2.92e-06  &    $-$1.41e-10  &    $-$1.06e-06  &    $-$3.21e-10  &    9.02e-11  \\
    J0900  &    0.0170  &    $-$2.36e-06  &    4.74e-10  &    $-$7.06e-06  &    $-$2.15e-10  &    7.36e-10  \\
    J0910  &    $-$0.0010  &    3.44e-06  &    $-$3.03e-10  &    $-$1.11e-06  &    $-$3.00e-10  &    1.57e-10  \\
    J1007  &    0.0040  &    4.35e-06  &    $-$4.19e-10  &    $-$4.36e-06  &    $-$1.90e-10  &    4.07e-10  \\
    uJPAS  &    $-$0.0638  &    8.69e-06  &    $-$4.91e-10  &    1.15e-05  &    6.44e-11  &    $-$6.18e-10  \\
    gSDSS  &    0.0012  &    $-$7.54e-07  &    3.20e-11  &    $-$3.00e-07  &    8.31e-11  &    4.43e-11  \\
    rSDSS  &    0.0000  &    0.00e+00  &    0.00e+00  &    0.00e+00  &    0.00e+00  &    0.00e+00  \\
    iSDSS  &    0.0012  &    $-$7.54e-07  &    3.20e-11  &    $-$3.00e-07  &    8.31e-11  &    4.43e-11  \\
    \hline
    \end{tabular}
  %\begin{tablenotes}
  %\item[a] 
  $f(x,y)=a_0+a_1y+a_2y^2+a_3x+a_4xy+a_5x^2$, where $x$ and $y$ denote CCD positions. $Mag_{new} = Mag + f(x,y)$.
  %\end{tablenotes}
  \end{threeparttable}
    \end{table*}

\begin{table*}
    \centering
    \caption{Fitting coefficients of metallicity-dependent stellar loci 
    for FGK dwarf stars in different colours.}
    \begin{threeparttable}
    \begin{tabular}{crrrrrr}
    \hline
    Coeff.  &    $uJAVA-r$  &    $J0378-r$  &    $J0390-r$  &    $J0400-r$  &    $uJPAS-r$  \\
    \hline
    $a_0$  &    1.3497  &    0.7578  &    0.4611  &    0.2539  &    0.8639  \\
    $a_1$  &    0.1453  &    0.1115  &    0.4240  &    0.1342  &    0.0569  \\
    $a_2$  &    0.0542  &    0.1315  &    0.1777  &    $-$0.0088  &    0.0107  \\
    $a_3$  &    0.0005  &    0.0155  &    0.0037  &    $-$0.0166  &    $-$0.0182  \\
    $a_4$  &    $-$0.6589  &    0.0724  &    1.1356  &    1.2538  &    0.5197  \\
    $a_5$  &    0.3754  &    0.7846  &    $-$0.0411  &    0.0147  &    0.6201  \\
    $a_6$  &    0.0078  &    0.0104  &    $-$0.1737  &    $-$0.0198  &    0.0094  \\
    $a_7$  &    2.9279  &    2.9310  &    1.4789  &    0.4866  &    1.9751  \\
    $a_8$  &    $-$0.1754  &    $-$0.3882  &    $-$0.1480  &    $-$0.0474  &    $-$0.3577  \\
    $a_9$  &    $-$0.9773  &    $-$1.1619  &    $-$0.6525  &    $-$0.1882  &    $-$0.7711  \\
    \hline
    \end{tabular}
     \begin{tablenotes}
  
  \item
  $f(x,y)=a_0+a_1y+a_2y^2+a_3y^3+a_4x+a_5xy+a_6xy^2+ $ 
             $a_7x^2+a_8yx^2+a_9x^3$, where $x$ $\equiv$ $g-i$ and $y$ $\equiv$ \feh. 
  \end{tablenotes}
   \end{threeparttable}
    \end{table*}

\begin{table*}
    \centering
    \caption{Fitting coefficients of metallicity-dependent stellar loci for FGK dwarf stars in different colours.}
     \begin{threeparttable}
    \begin{tabular}{lrrrrrr}
    \hline
    colours  & $a_0$ &  $a_1$ &  $a_2$ &  $a_3$ & $a_4$ & $a_5$\\
    \hline 	
    $J0410-r$  &    0.1153  &    0.0905  &    0.0425  &    1.2298  &    0.0706  &    0.0987  \\
    $J0420-r$  &    $-$0.0064  &    0.0735  &    0.0392  &    1.4127  &    0.0837  &    0.0037  \\
    $J0430-r$  &    0.0831  &    0.0990  &    0.0084  &    1.2993  &    $-$0.0355  &    $-$0.0851  \\
    $J0440-r$  &    0.0802  &    0.0747  &    0.0162  &    1.0258  &    $-$0.0198  &    $-$0.0621  \\
    $J0450-r$  &    0.0372  &    $-$0.0093  &    0.0007  &    0.8531  &    0.0376  &    $-$0.0543  \\
    $J0460-r$  &    0.0531  &    0.0315  &    0.0116  &    0.7094  &    $-$0.0113  &    $-$0.0551  \\
    $J0470-r$  &    0.0374  &    0.0508  &    0.0225  &    0.6404  &    $-$0.0122  &    $-$0.0336  \\
    $J0480-r$  &    0.1181  &    0.0050  &    0.0044  &    0.3398  &    $-$0.0054  &    0.1350  \\
    $J0490-r$  &    0.1000  &    0.0166  &    0.0098  &    0.3847  &    0.0033  &    0.0900  \\
    $J0500-r$  &    0.0339  &    0.0080  &    0.0038  &    0.3861  &    0.0079  &    0.1357  \\
    $J0510-r$  &    0.0180  &    0.0019  &    0.0030  &    0.3172  &    0.0157  &    0.2427  \\
    $J0520-r$  &    $-$0.0879  &    $-$0.0276  &    $-$0.0025  &    0.5926  &    0.0528  &    0.0653  \\
    $J0530-r$  &    $-$0.0097  &    $-$0.0079  &    0.0000  &    0.3320  &    0.0176  &    0.0521  \\
    $J0540-r$  &    $-$0.0265  &    $-$0.0144  &    $-$0.0008  &    0.3178  &    0.0187  &    0.0238  \\
    $J0550-r$  &    $-$0.0184  &    0.0019  &    0.0032  &    0.2551  &    0.0005  &    0.0115  \\
    $J0560-r$  &    $-$0.0122  &    0.0068  &    0.0057  &    0.2237  &    0.0046  &    $-$0.0029  \\
    $J0570-r$  &    $-$0.0223  &    0.0057  &    0.0029  &    0.1810  &    $-$0.0052  &    $-$0.0154  \\
    $J0580-r$  &    $-$0.0234  &    $-$0.0101  &    $-$0.0008  &    0.1296  &    $-$0.0026  &    $-$0.0135  \\
    $J0590-r$  &    $-$0.0205  &    $-$0.0188  &    $-$0.0019  &    0.1016  &    0.0091  &    0.0093  \\
    $J0600-r$  &    $-$0.0264  &    $-$0.0133  &    $-$0.0039  &    0.0814  &    $-$0.0032  &    $-$0.0225  \\
    $J0610-r$  &    $-$0.0132  &    $-$0.0078  &    $-$0.0026  &    0.0566  &    0.0033  &    $-$0.0109  \\
    $J0620-r$  &    0.0038  &    0.0077  &    0.0014  &    0.0079  &    $-$0.0060  &    0.0145  \\
    $J0630-r$  &    $-$0.0049  &    $-$0.0046  &    $-$0.0019  &    $-$0.0193  &    0.0075  &    0.0152  \\
    $J0640-r$  &    $-$0.0128  &    0.0014  &    0.0013  &    $-$0.0216  &    0.0035  &    $-$0.0054  \\
    $J0650-r$  &    0.0281  &    $-$0.0014  &    $-$0.0004  &    $-$0.0789  &    0.0016  &    $-$0.0029  \\
    $J0660-r$  &    0.0675  &    0.0022  &    0.0018  &    $-$0.1267  &    0.0001  &    0.0002  \\
    $J0670-r$  &    $-$0.0072  &    0.0015  &    $-$0.0004  &    $-$0.1183  &    $-$0.0004  &    0.0046  \\
    $J0680-r$  &    0.0152  &    0.0048  &    $-$0.0011  &    $-$0.1994  &    $-$0.0043  &    0.0476  \\
    $J0690-r$  &    0.0142  &    0.0074  &    0.0010  &    $-$0.2188  &    $-$0.0009  &    0.0521  \\
    $J0700-r$  &    0.0077  &    0.0164  &    0.0047  &    $-$0.1811  &    $-$0.0014  &    0.0093  \\
    $J0710-r$  &    0.0149  &    0.0167  &    0.0049  &    $-$0.2047  &    0.0029  &    0.0115  \\
    $J0720-r$  &    0.0206  &    0.0410  &    0.0123  &    $-$0.2286  &    $-$0.0083  &    0.0172  \\
    $J0730-r$  &    0.0085  &    0.0197  &    0.0060  &    $-$0.2146  &    0.0092  &    $-$0.0070  \\
    $J0740-r$  &    0.0071  &    0.0273  &    0.0067  &    $-$0.2088  &    0.0004  &    $-$0.0243  \\
    $J0750-r$  &    0.0248  &    0.0402  &    0.0109  &    $-$0.2444  &    $-$0.0117  &    $-$0.0187  \\
    $J0760-r$  &    0.0236  &    0.0275  &    0.0024  &    $-$0.2567  &    $-$0.0097  &    $-$0.0220  \\
    $J0770-r$  &    0.0307  &    0.0273  &    0.0030  &    $-$0.2818  &    $-$0.0068  &    $-$0.0165  \\
    $J0780-r$  &    0.0345  &    0.0273  &    0.0047  &    $-$0.3148  &    $-$0.0096  &    $-$0.0112  \\
    $J0790-r$  &    0.0337  &    0.0434  &    0.0115  &    $-$0.3238  &    $-$0.0087  &    $-$0.0115  \\
    $J0800-r$  &    0.0359  &    0.0402  &    0.0118  &    $-$0.3174  &    $-$0.0062  &    $-$0.0225  \\
    $J0810-r$  &    0.0576  &    0.0519  &    0.0178  &    $-$0.4069  &    $-$0.0082  &    0.0184  \\
    $J0820-r$  &    0.0783  &    0.0477  &    0.0120  &    $-$0.4141  &    $-$0.0037  &    0.0144  \\
    $J0830-r$  &    0.0717  &    0.0423  &    0.0139  &    $-$0.4116  &    0.0066  &    0.0090  \\
    $J0840-r$  &    0.0824  &    0.0453  &    0.0109  &    $-$0.4290  &    $-$0.0082  &    0.0133  \\
    $J0850-r$  &    0.1161  &    0.0541  &    0.0084  &    $-$0.4312  &    $-$0.0109  &    0.0061  \\
    $J0860-r$  &    0.1194  &    0.0605  &    0.0100  &    $-$0.4315  &    $-$0.0177  &    $-$0.0046  \\
    $J0870-r$  &    0.1207  &    0.0657  &    0.0100  &    $-$0.4755  &    $-$0.0226  &    0.0141  \\
    $J0880-r$  &    0.0973  &    0.0354  &    0.0058  &    $-$0.4718  &    0.0018  &    0.0095  \\
    $J0890-r$  &    0.1247  &    0.0537  &    0.0131  &    $-$0.5238  &    $-$0.0112  &    0.0216  \\
    $J0900-r$  &    0.1533  &    0.0524  &    0.0133  &    $-$0.5697  &    $-$0.0139  &    0.0417  \\
    $J0910-r$  &    0.1343  &    0.0684  &    0.0124  &    $-$0.5601  &    $-$0.0278  &    0.0288  \\
    $J1007-r$  &    0.2354  &    0.0638  &    0.0173  &    $-$0.6406  &    $-$0.0097  &    0.0526  \\
    $gSDSS-r$  &    0.0357  &    0.0291  &    0.0069  &    0.7234  &    $-$0.0001  &    $-$0.0132  \\
  %  $rSDSS-r$  &    0.0000  &    0.0000  &    0.0000  &    0.0000  &    0.0000  &    0.0000  \\
    $iSDSS-r$  &    0.0357  &    0.0291  &    0.0069  &    $-$0.2766  &    $-$0.0001  &    $-$0.0132  \\
    \hline
    \end{tabular}
    \begin{tablenotes}
  \item $f(x,y)=a_0+a_1y+a_2y^2+a_3x+a_4xy+a_5x^2$, where $x$ $\equiv$ $g-i$ and $y$ $\equiv$ \feh.
  \end{tablenotes}
   \end{threeparttable}
    \end{table*}

Using the 310 miniJPAS-JPLUS FGK dwarf sample above, we have carried out a global two-dimensional polynomial fit to a series of 59 miniJPAS colours ($(X-r)_0$, where $X$ refers to one of the 56+4 miniJPAS filters, except for $r$) as a function of the $(g-i)_0$ colour and metallicity [Fe/H]. A few stars with photometric errors larger than 0.05\,mag are excluded from this step. Note that all colours referred to hereafter are dereddened intrinsic values, with reddening from \citet{Schlegel1998} and reddening 
coefficients from the \citet{Schlafly2016} extinction curve\footnote{The numbers are stored in the miniJPAS.Filter table, available online.}. 
For colours $uJAVA - r$, $J0378-r$, $J0390-r$, $J0400-r$, and $uJPAS-r$, a third-order polynomial with 10 free parameters is adopted. For the remaining colours that are relatively insensitive to metallicity, 
a second-order polynomial of 6 free parameters is used to avoid over-fitting. A two-sigma clipping is performed during the fitting process. 

\begin{figure*}
	\centering
	\includegraphics[width=\textwidth]{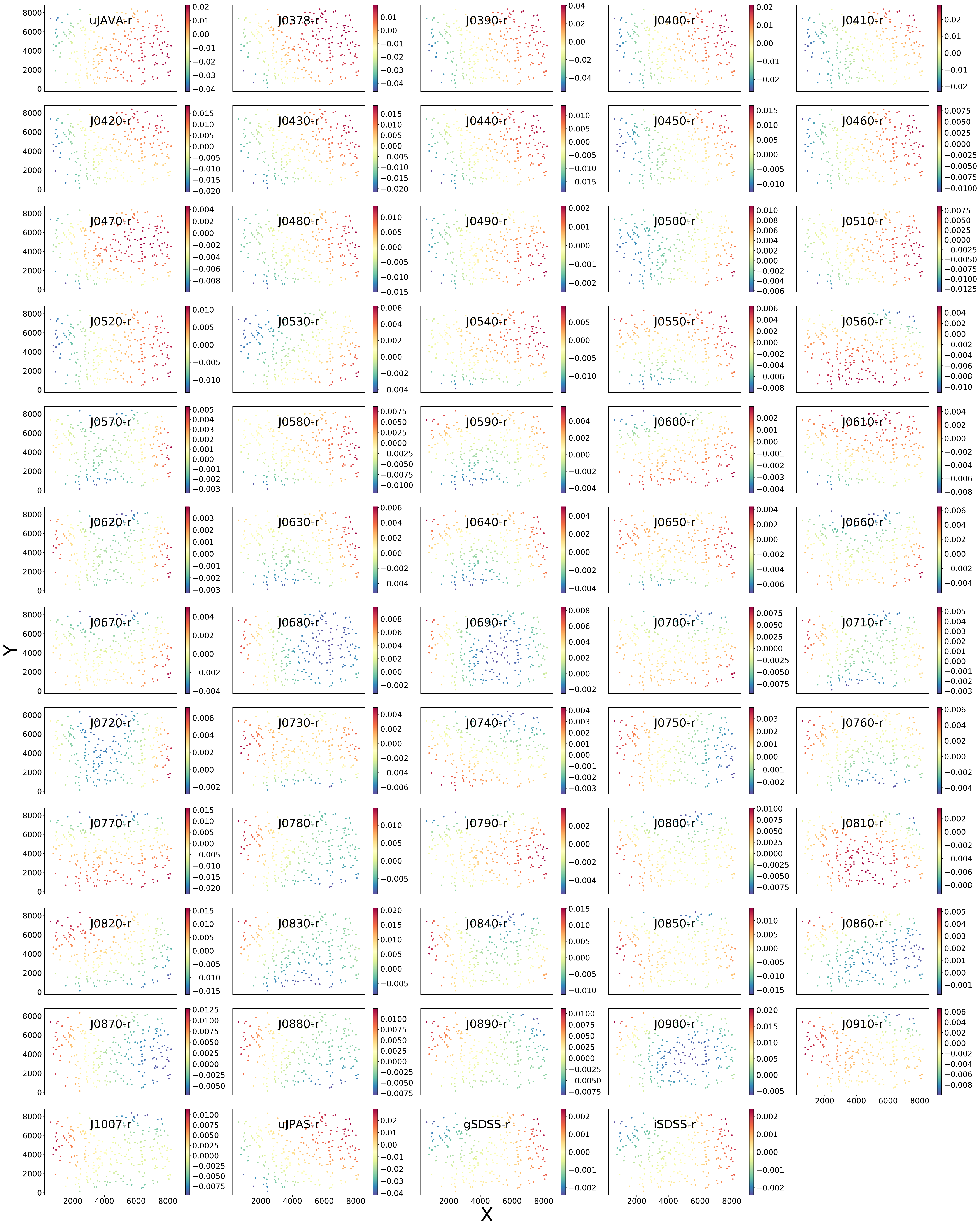}
	\caption{Flat-fielding errors in units of magnitude in the CCD focal plane for different miniJPAS colours (see text). The errors are derived by a 2D polynomial fitting as a function of CCD (X, Y) positions to the locus residuals. Note: the scales are different for different colours.  Note: $r$ $\equiv$ $rSDSS$.}
	\label{fig:locus}
\end{figure*}

\begin{figure*}
	\centering
	\includegraphics[width=\textwidth]{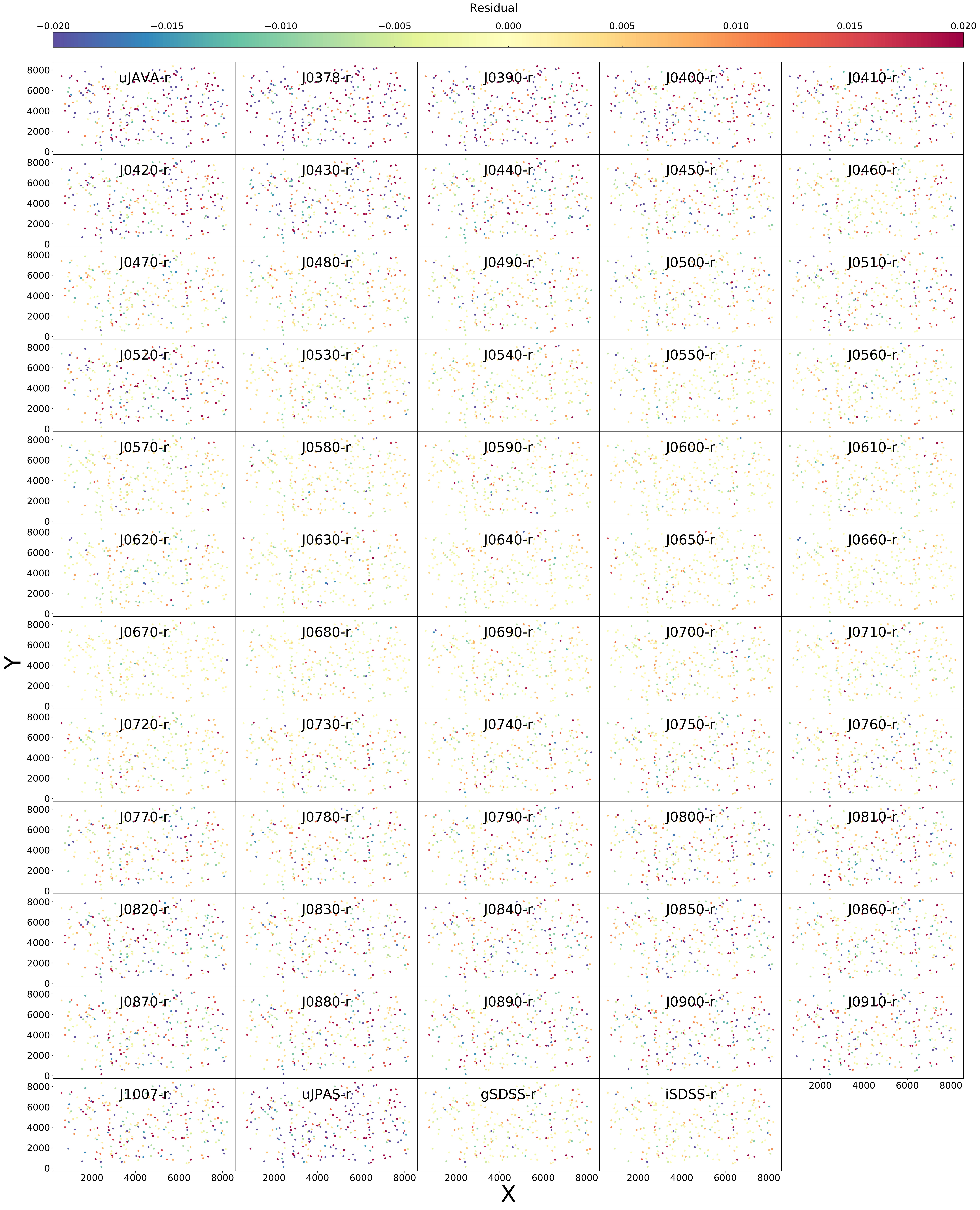}
	\caption{Fitting residuals in the CCD focal plane after correcting for the derived flat-fielding errors. No spatial patterns are found.}
	\label{fig:flat}
\end{figure*}

\begin{figure*}
	\centering
	\includegraphics[width=\textwidth]{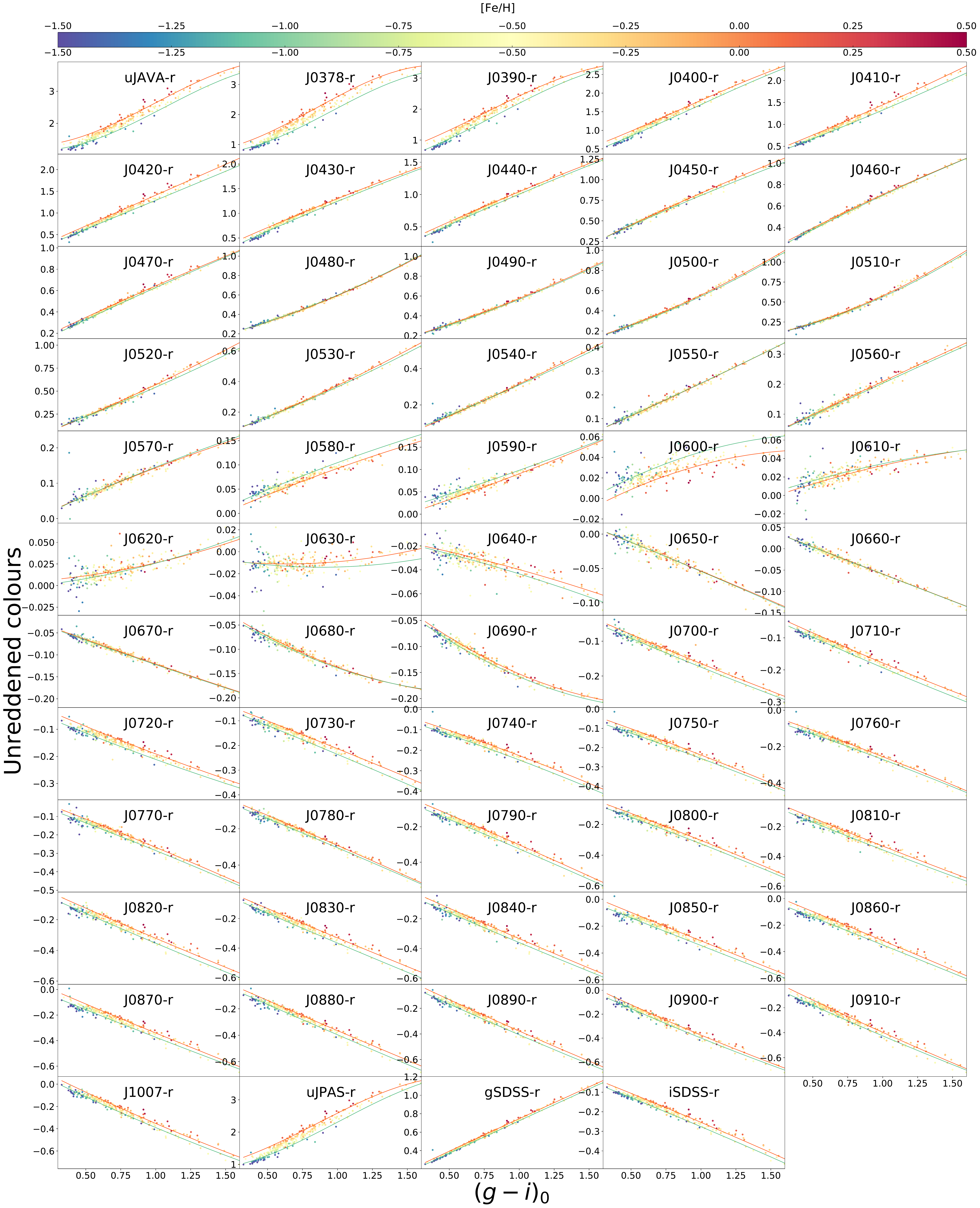}
	\caption{Metallicity-dependent stellar loci of different miniJPAS filters. In each panel, the dwarf stars are colour coded by their metallicities. The red and cyan lines, predicted by the polynomials of Table\,2 and Table\,3, correspond to [Fe/H] of 0.0 and $-$1.0, respectively.}
	\label{fig:flatresi}
\end{figure*}

\begin{figure}
	\centering
	\includegraphics[width=\columnwidth]{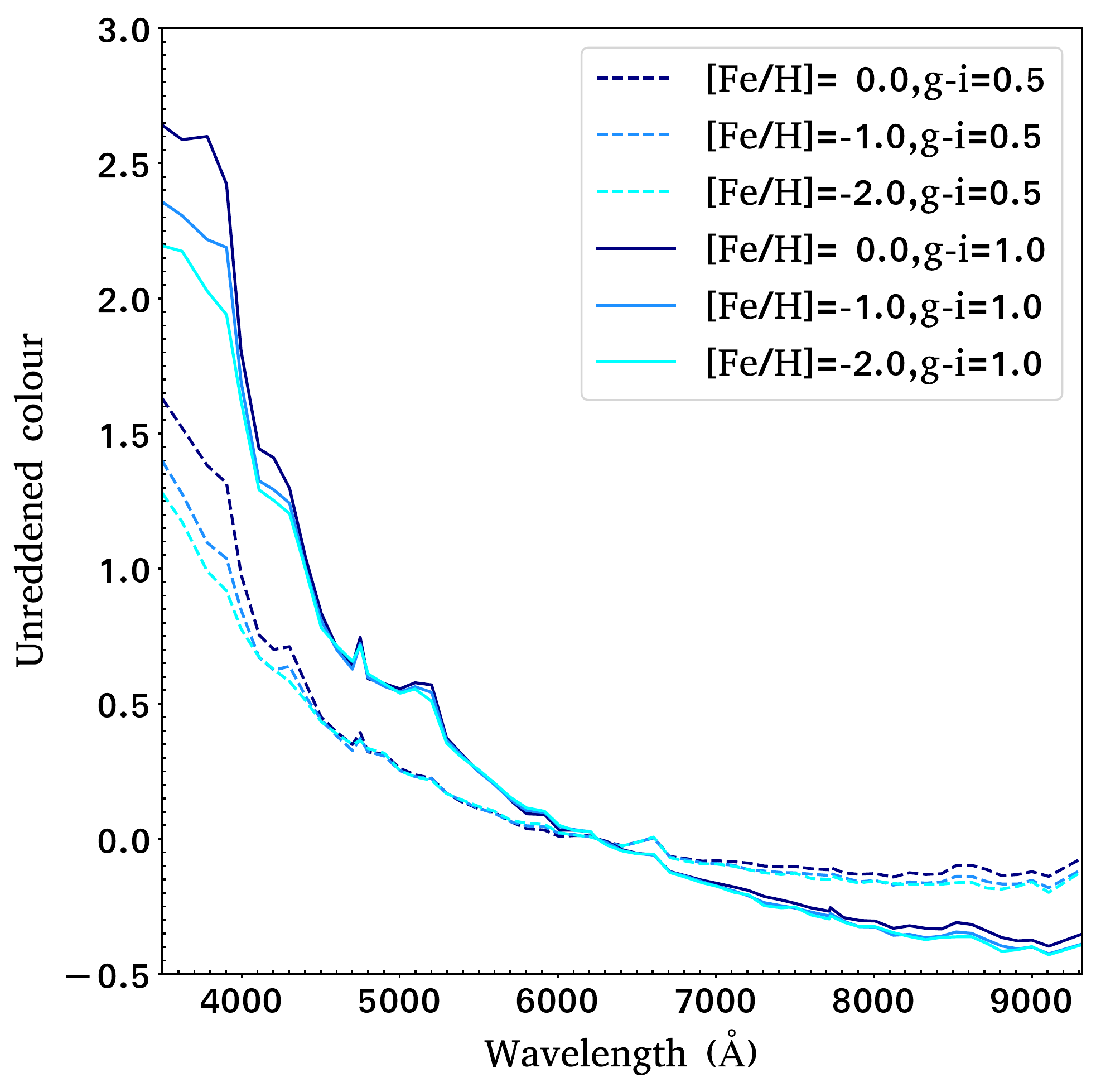}
	\caption{Comparison of $X-r$ colours between stars of the same $g-i$ colour (dashed lines: $g-i=0.5$, solid lines: $g-i=1.0$) but with different metallicities, where $X$ refers to a given miniJPAS filter. }
	\label{fig:c_wave}
\end{figure}

The fitting residuals show clear spatial variations in the CCD focal plane, from about 0.2 per cent in the red colours to a few per cent in the blue colours.  These spatial variations are probably caused by flat-fielding (illumination correction) errors in the miniJPAS data. Therefore, we have performed a 2D polynomial fitting as a function of CCD (X, Y) positions to the initial locus residuals. The fitting coefficients are listed in Table\,1. The results of flat-fielding errors are displayed in Figure\,\ref{fig:locus}. The fitting residuals after correction are displayed in Figure\,\ref{fig:flat}.  
In the next step, we have revised the miniJPAS colours/magnitudes accordingly to 
correct for their flat-fielding errors, assuming that there are no errors in the $r$ band\footnote{Note that the revised magnitudes have also been used in the SED fittings with VOSA in Section\,3.}. The metallicity-dependent stellar loci of different miniJPAS filters are fitted again. The results are shown in 
Figure\,\ref{fig:flatresi}.
The resultant fitting coefficients are listed in Table\,2 and Table\,3.  The differences
between the initial fitting curves and those after one iteration are,
on average, smaller than 0.01\,mag, so no further iterations are needed.
Figure\,\ref{fig:c_wave} compares colours between stars of the same $g-i$ colour but with different metallicities, which are estimated using the fitted polynomials. The large differences in the blue filters are seen. 

\begin{figure}
%	\centering
	\includegraphics[width=\columnwidth]{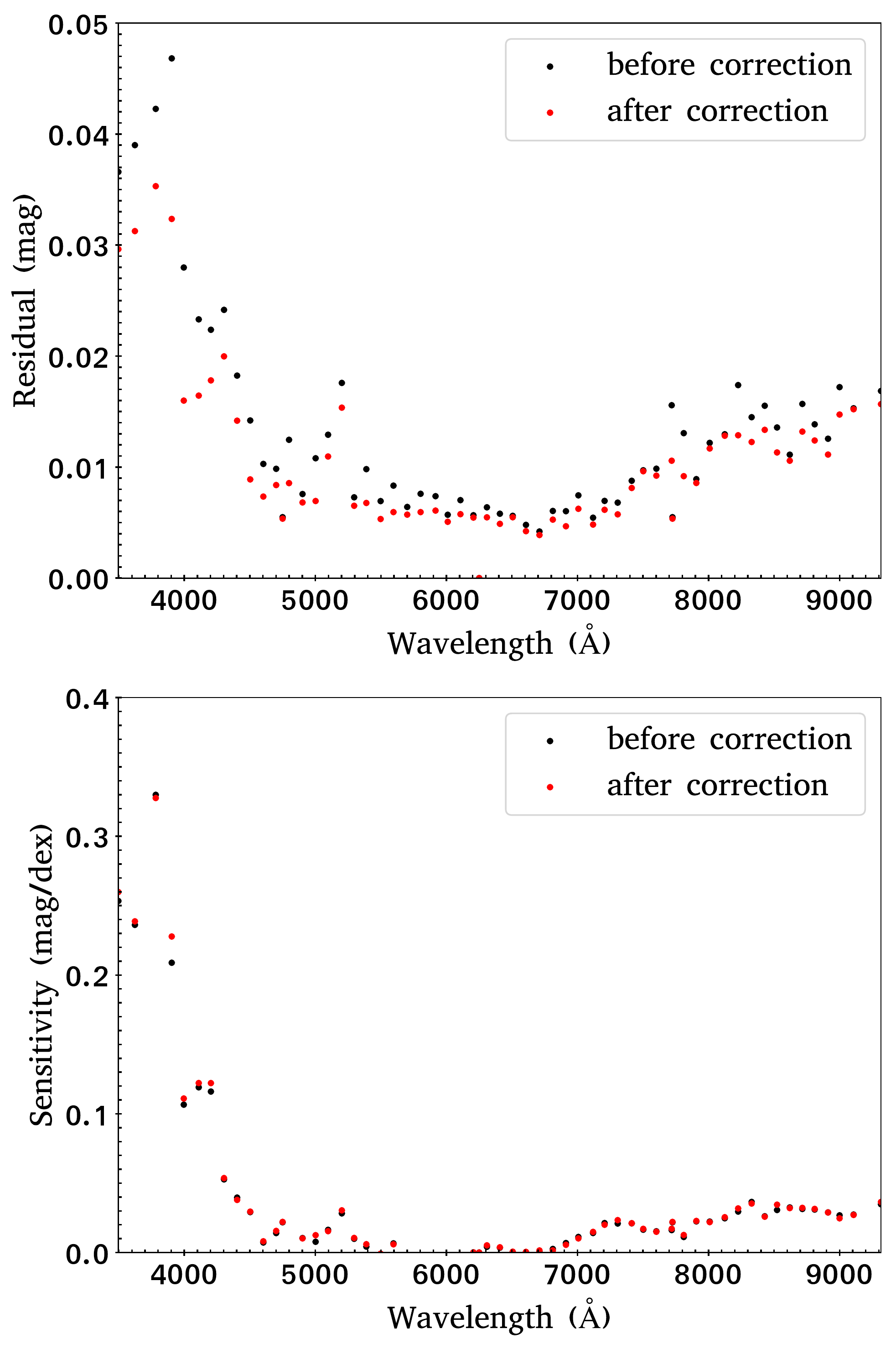}
	\caption{{\sl Top panel:} Fitting residuals of metallicity-dependent stellar loci for different $X-r$ colours. {\sl Bottom panel:} Typical metallicity sensitivities for different $X-r$ colours. The black and red dots represent results before and after the correction, respectively.}
	\label{fig:plot_locus_4paper}
\end{figure}

More details are plotted in Figure\,\ref{fig:plot_locus_4paper}.
The top panel of Figure\,\ref{fig:plot_locus_4paper} compares the locus fitting residuals before and after correction. It can be seen that the residuals decrease remarkably after correction, particularly for the blue colours. The fitting residuals are larger 
in the bluer colours for two main reasons: the stronger dependence on metallicity and the larger photometric errors. The larger numbers in the redder colours are due to similar reasons. The typical residuals are around 0.005\,mag, suggesting that the photometric quality of the miniJPAS data is very high. 
Note that there is a strong peak around the J0510 filter,  which is caused by the presence of magnesium absorption features. 
Due to the MgII triplet, the J0510 and J0520 filters show strong dependence on the Mg abundance (see \citealt{Yang2022} for more details).
Therefore, variations of [Mg/Fe] for a given [Fe/H] cause the larger fitting residuals. For similar reason, there is another peak in the J0430 filter, 
caused by the effect of [C/Fe] abundance and CH G-band feature. 
The peak at the J0390 filter is likely caused by the 
[Ca/Fe] abundance and CaII HK lines. 
The peak at the J0378 filter is probably due to its strongest metallicity dependence.
The peaks at the J0390, J0430, J0510, and J0520 filters suggest that
we can estimate individual element abundances such as [Ca/Fe], [C/Fe], and [Mg/Fe] from the miniJPAS/JPAS data, 
as already demonstrated with the JPLUS data by \citet{Yang2022}. 

The bottom panel of Figure\,\ref{fig:plot_locus_4paper} shows the typical metallicity sensitivities for different colours, which are defined as the median variation of a given colour 
loci of \feh~ = $-$1 and \feh~ = 0.
The very blue colours, including $uJAVA - r$, $J0378-r$, $J0390-r$, $uJPAS-r$,  show very strong metallicity dependence, around 0.25 mag/dex.  The sensitivities decrease to about 0.1 mag/dex for the $J400-r$, $J410-r$, and $J420-r$ colours, 
and then to 0.01 mag/dex. The numbers increase to 0.03 mag/dex for colours redder than 7000\,\AA.

\section{Photometric metallicites of FGK dwarf stars from miniJPAS} \label{feh}

In this section, we combine the empirical metallicity-dependent stellar loci of section\,\ref{sl} and a minimum  $\chi^2$ technique to determine photometric metallicities of FGK dwarf stars from the miniJPAS data. Similar approaches have been applied to the SDSS/Stripe 82 \citep{Yuan2015b,Zhang2021}, {\it Gaia} EDR3 \citep{Xu2022}, and SMSS DR2 \citep{Huang2022}. 
Here the $\chi^2$ is defined as:

\begin{equation}
%\begin{aligned}
   \chi^2(\feh, g-i) = 
   \sum_{i=1}^{59}\frac{[c_{\rm obs}^i - R_{c}^{i} \times E(B-V) - c_{\rm int}^i(\feh, g-i) ]^2}{(\sigma_{c}^{i})^2 \times (59-2)},
%\end{aligned}
\end{equation}
\\
where $c_{\rm obs}^i$ ($i$ = 1 -- 59) are the observed miniJPAS colours.
$R_{c}^{i}$, $c^i_{\rm int}$([Fe/H], $g-i$), and $\sigma^i_c$  (i = 1 -- 59) are respectively  
the reddening coefficients, the intrinsic colours predicted by the metallicity-dependent stellar loci for a given set of metallicity \feh~and 
intrinsic colour $g-i$, and the uncertainties of observed colours.
Values of \ebv~are from the extinction map of \citet{Schlegel1998}. 
Values of $\sigma_i$ are estimated from the magnitude errors and calibration uncertainties. 
The calibration uncertainties are adopted as 
0.005 mag for all colours. The number is chosen for two reasons: 1) 
it is the typical precision we can achieve for blue filters with our method (e.g., \citealt{HuangYuan2022}); 2) it is very close to the minimum value 
of the fitting residuals after correction in Figure\,\ref{fig:plot_locus_4paper}. The calibration uncertainties are likely over-estimated for filters around the $r$ band, but it will not affect the main result of this work, as the calibration uncertainties are usually much smaller than 
the magnitude errors.
 
We use a brute-force algorithm to determine the optimal \feh~for each sample star.
For a given sample star, the value of \feh~is varied from $-2.5$ to 0.5 at a step of 0.01\,dex.
A 1D array of 301 $\chi^2$ values is 
calculated to find the minimum value of $\chi^2$,
$\chi^2_{\rm min}$, and the corresponding metallicity \feh. 
A Monte Carlo method is used to estimate the errors of \feh~
by introducing a Gaussian random noise for each colour 200 times.  

\begin{figure*}
	\centering
	\includegraphics[width=\textwidth]{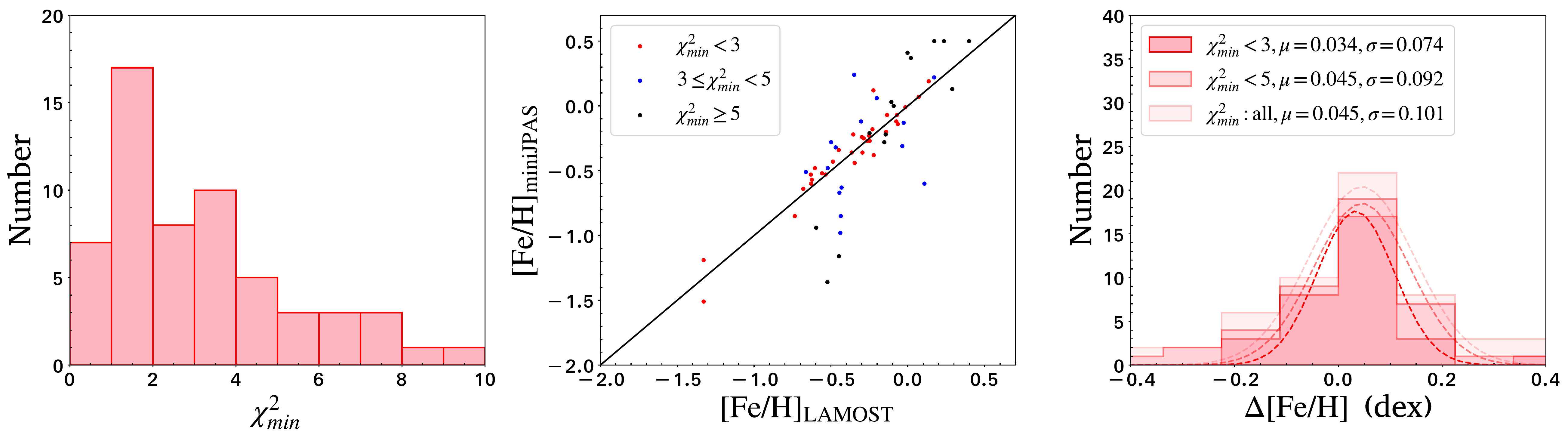}
	\caption{Left: histogram distribution of the minimum $\chi^2$ for the LAMOST-miniJPAS test sample. Middle: Comparison between miniJPAS and LAMOST metallicities. The black line is the identify function. Right: histogram distributions of the metallicity differences. The Gaussion fitting results are overplotted, with the mean and sigma values labelled.}
	\label{fig:check_feh_4paper}
\end{figure*}

\begin{figure}
	\centering
	\includegraphics[width=1.0\columnwidth]{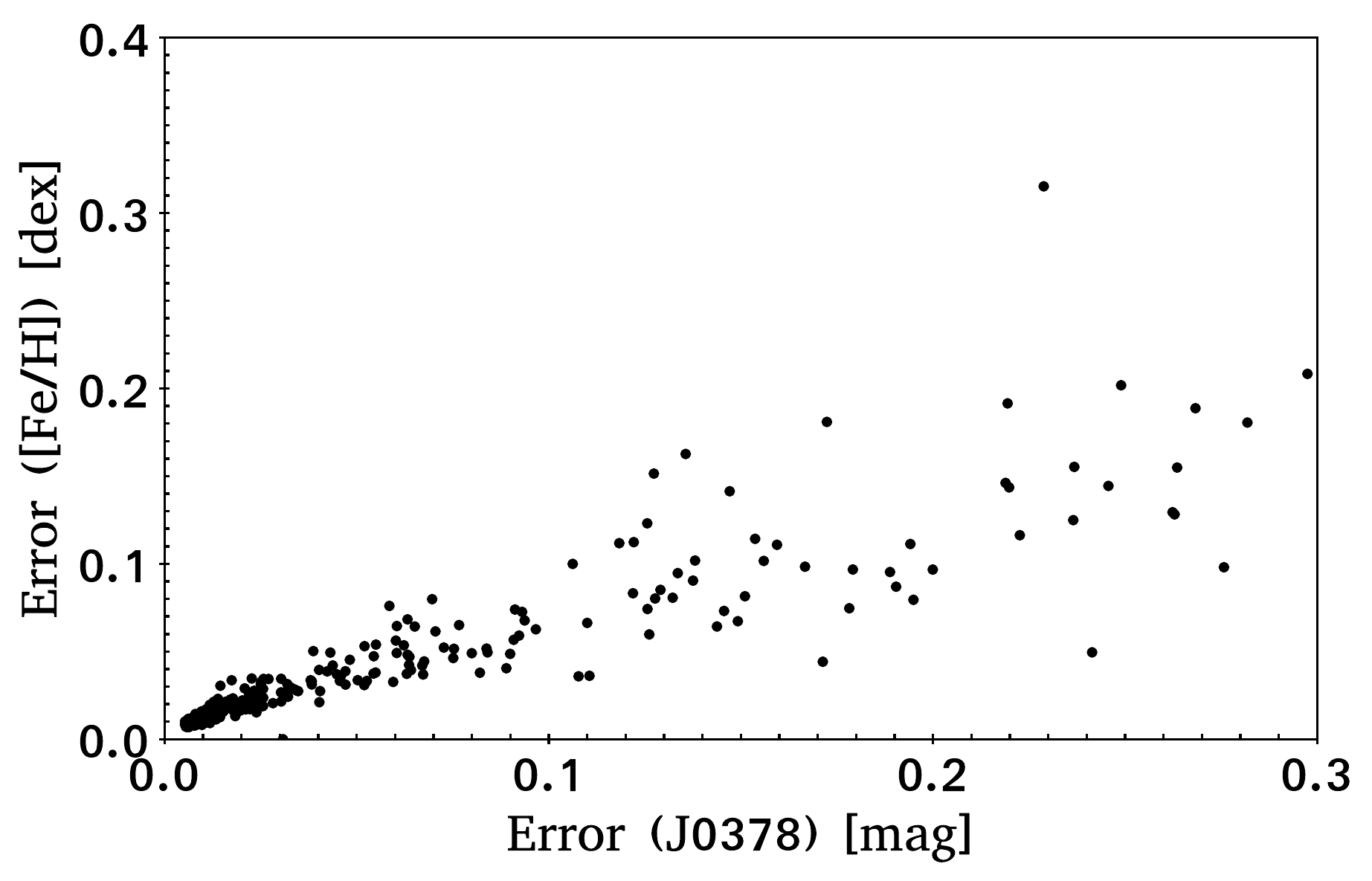}
	\caption{ [Fe/H] errors as a function of J0378 errors for stars of [Fe/H] > $-$1. }
	\label{fig:feherr_vs_err}
\end{figure}

We apply the method to 683 FGK stars of high quality photometry ($0.3 < g-i < 1.6$, flag = 0 in all filters, $rmag < 20$), assuming that they are all dwarfs\footnote{Note that we expect a few per cent giant stars
in the sample. In this work,  we did not obtain metallicity-dependent        
stellar loci for giant stars due to the lack of sufficient sample. 
Stellar loci are usually different between dwarfs and giants \citep[e.g.,][]{Zhang2021}.
With up-coming JPAS data, the differences can be used to discriminate giants 
from dwarfs, as we did to the SDSS/Stripe 82 \citep{Zhang2021} }.
The median of the $\chi^2_{\rm min}$ values is 1.4. The typical 
$\chi^2_{\rm min}$ values are slightly larger when the stars are brighter, suggesting that the uncertainties are slightly under-estimated for bright stars.
To test the precision of the method, we have selected 61 dwarf stars in common with LAMOST DR7 and having LAMOST metallicity uncertainties smaller than 0.15\,dex.
The left panel of Figure\,\ref{fig:check_feh_4paper} shows the histogram distribution of the $\chi^2_{\rm min}$ values for the LAMOST-miniJPAS test sample. The median value is about 2.6. 
The middle and right panels compare the miniJPAS and LAMOST metallicities. 
A good agreement is seen, particularly for those of $\chi^2_{\rm min} < 3$. 
The metallicity differences have a mean value close to zero and a sigma value of about 0.1\,dex, suggesting that the errors of our photometric metallicities are smaller than 0.1\,dex. 
The results are not surprising, given the strong dependence on metallicity of the blue miniJPAS filters and the high-quality of the data.
Note in the middle panel of Figure\,\ref{fig:check_feh_4paper}, there are several stars
having miniJPAS metallicities lower than LAMOST ones by 0.3 dex and $\chi^2_{\rm min} > 3$. They are 
probably binaries, as metallicities could be under-estimated significantly for certain binaries by our method \citep[e.g.,][]{Yuan2015b,Xu2022}.
One would expect that the \feh~errors depend on photometric errors, stellar colour, and \feh.
Figure\,\ref{fig:feherr_vs_err} plots 
\feh~errors estimated from Monte Carlo simulations as a function of errors in the $J0378$ band for stars of [Fe/H] > $-$1.
A tight linear relation is found. Similar relations are found for 
other filters, but with steeper slopes for filters that are less sensitive to \feh. 
The results demonstrate the power and potential of miniJPAS/JPAS in precise metallicity estimates for a large number of stars.

\begin{figure}
	\centering
	\includegraphics[width=1.0\columnwidth]{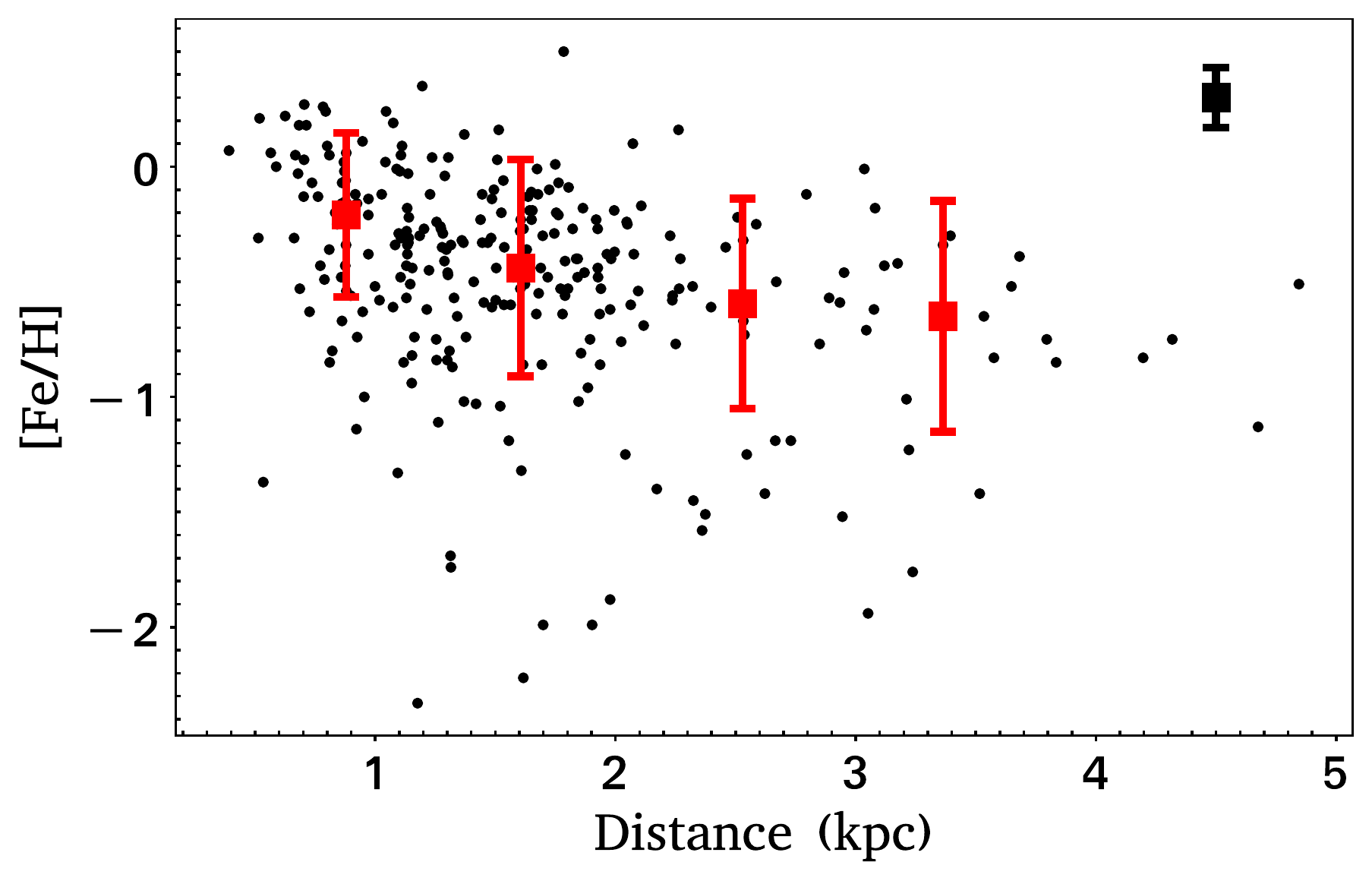}
	\caption{[Fe/H] as a function of distance for 269 selected dwarfs. See more details in Section\,5. The red dots with error bars denote the median values and standard deviations of the black dots. The black error bar denotes the typical error of  $\sigma$[Fe/H] $=$ 0.13\,dex.}
	\label{fig:feh_vs_distance}
\end{figure}

We cross-match the 683 stars with the {\it Gaia} EDR3 and further select dwarf stars ($M_r>10*(g-i)-4.0$ or $M_r>4.0$) of $\chi^2_{\rm min}$ < 5, PARALLAX > 0, and PARALLAX\_OVER\_ERROR $>$ 5, where $M_r$ is absolute magnitude in the $r$ band.
Figure\,\ref{fig:feh_vs_distance} plots metallicites of the selected 269 dwarf stars as a function of distance. A clear metallicity gradient is found, as expected. Note that stars of [Fe/H] $< -1.5$ may suffer large 
uncertainties due to the loci of $< -1.5$ are not well constrained yet.
There are a few stars with distances less than 1.5\,kpc and [Fe/H] $< -1.5$. They are more likely binary stars in the disk rather than metal-poor stars in the halo, for
reasons mentioned above that metallicities could be under-estimated significantly for binaries.
Besides, most of them have tangential velocities similar to those of disk 
stars (Xu et al., 2022). %add reference

\section{Giant dwarf classification}\label{giant}

Stellar loci between dwarf stars and giant stars are different, particularly in the blue filters \citep[e.g.,][]{Zhang2021} and certain narrow-band filters \citep[e.g.,][]{Morrison2000}. However, due to the 
very limited giant stars available in the miniJPAS data, it is very
challenging to determine metallicity-dependent stellar loci of giant stars. 
With the help of {\it Gaia} parallaxes, we are able to explore the potential of 
giant/dwarf classification with the 60 miniJPAS filters in this section.

The miniJPAS stars are cross-matched with the {\it Gaia} EDR3.
In this work, we only use stars with high-precision photometry. The following criteria are required: 
1) Flag $=$ 0 in all miniJPAS filters; 2) ERR\_APER6 $<$ 0.25 in all miniJPAS filters; and 3) PARALLAX\_OVER\_ERROR $>$ 9.0 for dwarfs, $>$ 1.0 for giants due to their much larger distances and consequently larger relative parallax errors.

\begin{figure}
	\centering
	\includegraphics[width=1.0\columnwidth]{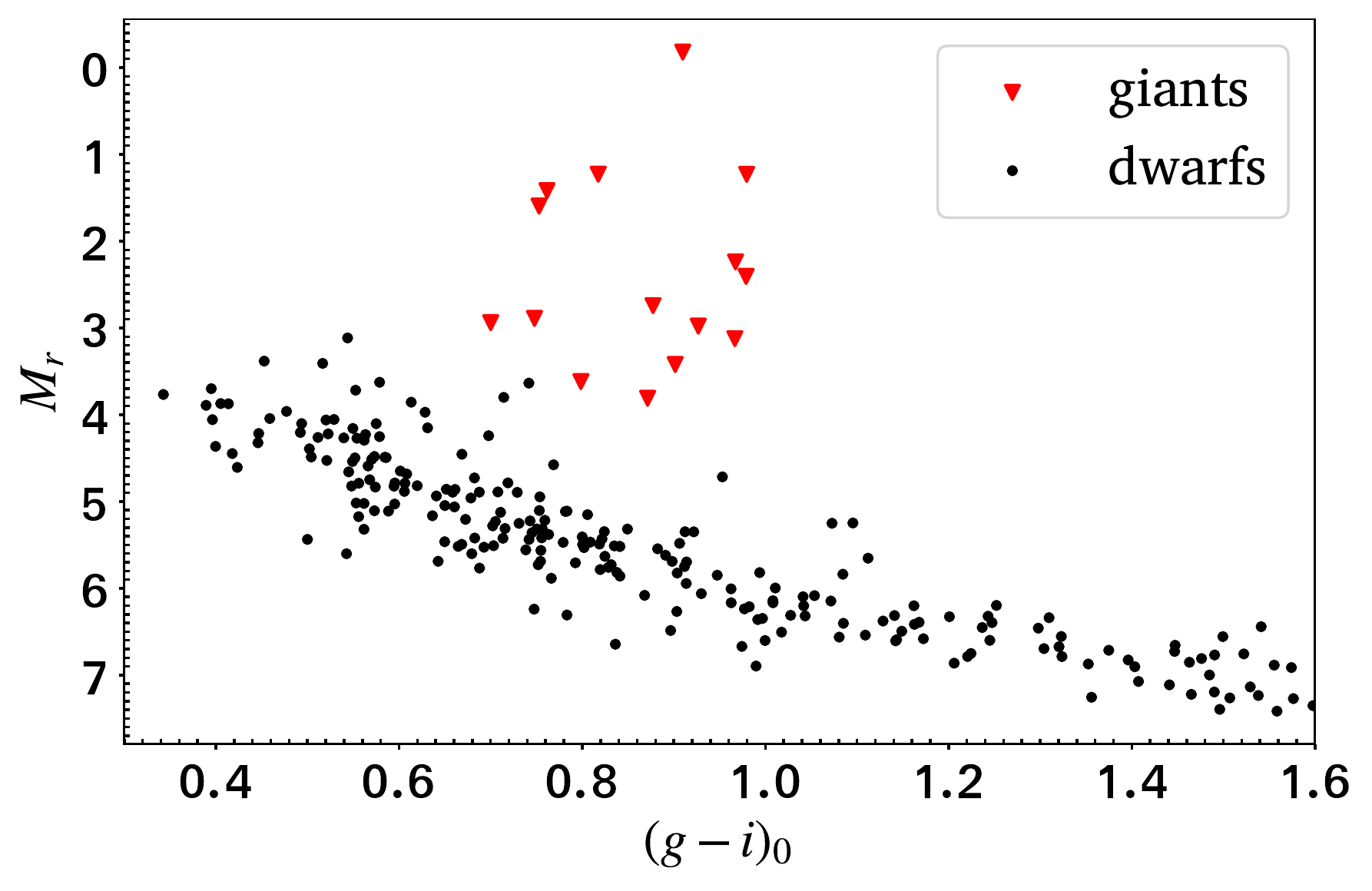}
	\caption{Distribution of the giant stars (red triangles) and dwarf stars (black dots) in the HR diagram.}
	\label{fig:HR}
\end{figure}

\begin{figure*}
	\centering
	\includegraphics[width=\textwidth]{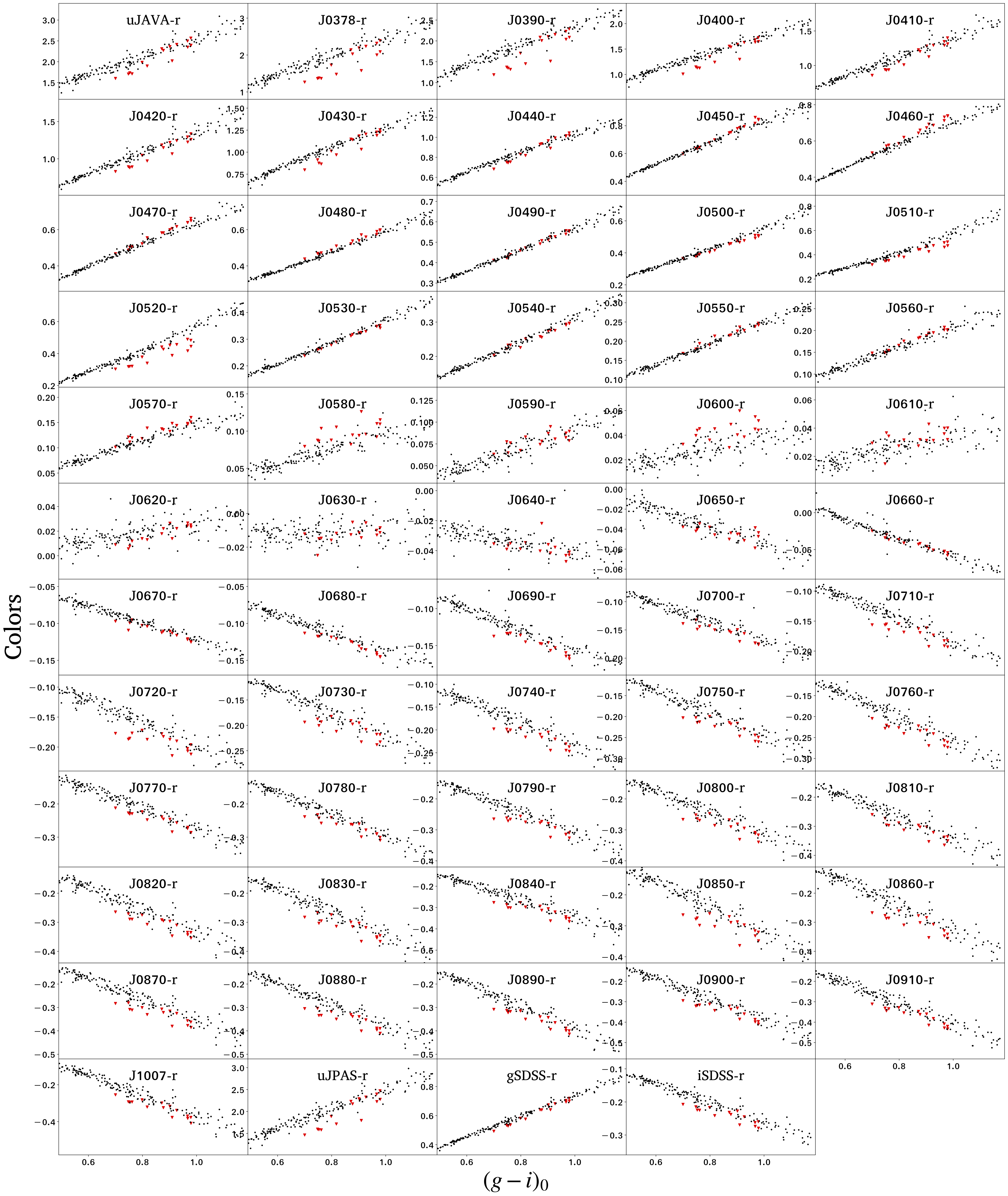}
	\caption{Distributions of the giant stars (red triangles) and dwarf stars (black dots) in the colour-colour diagrams. Note that the ranges in y-axes are different.}
	\label{fig:colour_vs_gi}
\end{figure*}

Dwarf stars are selected empirically as those of $M_r >10*(g-i)-4.0$ or $M_r>4.0$,
while giant stars of $M_r<10*(g-i)-4.0$ and $M_r<4.0$.
Finally we have obtained a sample of 233 dwarf stars and 15 giant stars. 
Their distribution in the HR diagram is shown in Figure\,\ref{fig:HR}.
Their distributions in different colour-colour diagrams are shown in Figure\,\ref{fig:colour_vs_gi}.

The extreme gradient boosting ({\it XGBoost}; \citealt{2016arXiv160302754C}) is chosen as our classification model for the giant/dwarf stars. This algorithm can select important features during training process and mitigate over-fitting by pruning, which improves performance of small sample classification. All the J-PAS filters are used. The training process is carried out with the python wrapper of XGBoost provided by scikit-learn. The best values of the hyperparameters are determined by the grid search with 5-fold cross validation, and values ('learning\_rate': 0.02, 'n\_estimators': 90, 'max\_depth': 4, 'min\_child\_weight': 1, 'seed': 0,'subsample': 0.9, 'colsample\_bytree': 0.6, 'gamma': 0.02, 'reg\_alpha': 0.00009, 'reg\_lambda': 0.0008) are used as the final hyperparameters in this paper. 

\begin{figure}
	\centering
	\includegraphics[width=1.0\columnwidth]{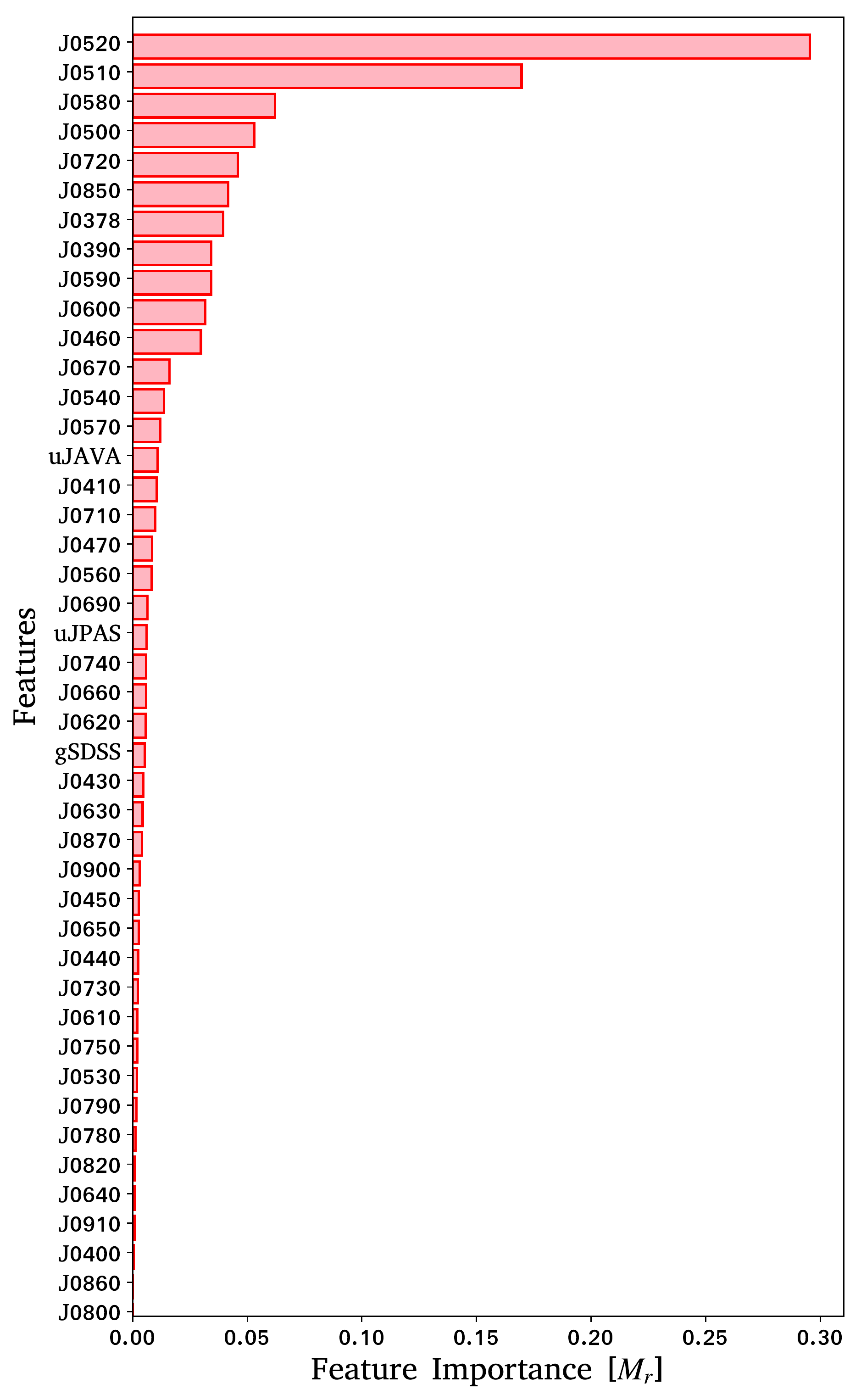}
	\caption{Relative importance of the features for the giant/dwarf classification.}
	\label{fig:feature_impor}
\end{figure}

The total samples are divided into the training set and the test set according to the ratio of 3:1. Both the training and testing sets are classified correctly, suggesting the strong potential of the miniJPAS filters in discriminating dwarfs and giants. We rank all the features (shown in Figure\,\ref{fig:feature_impor}) according to the total gain when the algorithm splits candidates.
It can be seen that J0520 and J0510 are the two most important filters, as expected, because the Mg triplet is more prominent in dwarfs than in giants for FGK stars. At the same $g-i$ colour, giants are usually bluer in 
$J0510-r$ and $J0520-r$ colours. Note that Mg abundances also show moderate 
effect on $J0510-r$ and $J0520-r$ colours \citep{Yang2022}. Note also that 
due to the limited number and parameter space (e.g., in \feh)  of both 
giants and dwarfs in the above classification,  the ranking of the features is 
probably biased. Further studies are needed to address this problem when 
the J-PAS data are available.

\section{Conclusions} \label{summary}
 
The unique set of 54 overlapping narrow band and 2 two broad filters of the J-PAS survey provides a great opportunity for stellar physics and Galactic studies.
In this work, we use the miniJPAS data to explore and prove the potential of the J-PAS filter system in characterizing stars and deriving their fundamental parameters via 
different techniques. The main conclusions are as follows.

   \begin{enumerate}
   
   \item We used VOSA to derive stellar effective temperatures via SED fitting using  miniJPAS photometry. We found that effective temperatures are in good agreement with those obtained from spectroscopic data from LAMOST and with previous works. We estimated a typical uncertainties $<$150\,K, which validate the high quality of the J-PAS data and postulates SED fitting technique as a suitable way to obtain reasonably accurate effective temperatures for the million of stars expected to be observed in the J-PAS survey.
      \item We have constructed the metallicity-dependent stellar loci in 59 colours for the miniJPAS FGK dwarf stars. The locus fitting residuals show significant 
      spatial variations in the CCD focal plane, from about 0.2 per cent in the red colours to a few per cent in the blue colours. Such patterns, probably due to errors in the flat-fielding process, have been corrected.  
      The very blue colours, including $uJAVA - r$, $J0378-r$, $J0390-r$, $uJPAS-r$,  show the strongest metallicity dependence, around 0.25 mag/dex.  The sensitivities decrease to about 0.1 mag/dex for the $J0400-r$, $J0410-r$, and $J0420-r$ colours.
    The fitting residuals show peaks at the $J0390, J0430, J0510$, and $J0520$ filters, suggesting that 
    we can estimate individual elemental abundances such as [Ca/Fe], [C/Fe], and [Mg/Fe] from the miniJPAS/J-PAS data, as already demonstrated with the JPLUS data by \citet{Yang2022}. 
      
      \item Combining the empirical metallicity-dependent stellar loci and a straight-forward 
      minimum  $\chi^2$
      technique, we have achieved a metallicity precision better than 0.1 dex solely from the miniJPAS photometry. The results demonstrate the power and potential of miniJPAS/J-PAS in precise metallicity estimates for a huge number of stars in the future. Given the strong dependence on metallicity of the blue miniJPAS/J-PAS filters and the high data quality, we expect to achieve a high precision
      for stars with a metallicity down to $\sim -3.5$ or lower in the future. 
      
      \item We have used {\it XGBoost} to classify dwarfs and giants. Both the training and testing samples are all successfully classified, suggesting the strong potential of the miniJPAS/J-PAS filters in discriminating dwarfs and giants. The $J0520$ and $J0510$ filters play an import role in the 
      classifications, as expected. 
   \end{enumerate}

The results obtained in this work  demonstrate the power and potential of miniJPAS/J-PAS data in stellar parameter determinations.
Due to the very small number of stars in the small miniJPAS footprint, 
scientific investigations are limited. 
Note a very nice exploration of white dwarf science using the miniJPAS data by  \citet{Lopez2022}.
In the future with the J-PAS data, stellar parameters can be better determined, not only for effective 
temperature, metallicity, and surface gravity, but 
also for reddening, distance, age, [C/Fe], [N/Fe], [Mg/Fe], and [Ca/Fe]. 
Such a magnitude-limited sample of stars down to $r \sim$ 21 -- 22 and within 8500 $deg^2$
will be valuable for studies of 
stellar populations, structure, chemistry, and evolution of the Galaxy. 

\section*{acknowledgements}
We acknowledge the referee for his/her valuable comments and suggestions that improved the quality of the paper significantly. 
This work is supported by the National Natural Science Foundation of China through the projects NSFC 12222301, 12173007, 11603002, National Key Basic R \& D Program of China via 2019YFA0405500, 
and Beijing Normal University grant No. 310232102. 
We acknowledge the science research grants from the China Manned Space Project with NO. CMS-CSST-2021-A08 and CMS-CSST-2021-A09. 
This research has made use of the Spanish Virtual Observatory (https://svo.cab.inta-csic.es) project funded by MCIN/AEI/10.13039/501100011033/ through grant PID2020-112949GB-I00.
PC acknowledges financial support from the Government of Comunidad Autónoma de Madrid (Spain), via postdoctoral grant ‘Atracción de Talento Investigador’ 2019-T2/TIC-14760. 
The work of V.M.P. is supported by NOIRLab, which is managed by the Association of Universities for Research in Astronomy (AURA) under a cooperative agreement with the National Science Foundation.
F.J.E. acknowledges financial support by the Spanish grant MDM-2017-0737 at Centro de Astrobiolog\'{\i}a (CSIC-INTA), Unidad de Excelencia Mar\'{\i}a de Maeztu.
C.A.G. awknowledges finantial support from the CAPES through scolarship for developing his Ph.D project and any related research.
Part of this work was supported by institutional research funding IUT40-2, JPUT907 and PRG1006 of the Estonian Ministry of Education and Research. We acknowledge the support by the Centre of Excellence “Dark side of the Universe” (TK133) financed by the European Union through the European Regional Development Fund.

Based on observations made with the JST/T250 telescope and JPCam at the Observatorio Astrofísico de Javalambre (OAJ), in Teruel, owned, managed, and operated by the Centro de Estudios de F{\'i}sica del Cosmos de Arag{\'o}n (CEFCA). 
We acknowledge the OAJ Data Processing
and Archiving Unit (UPAD) for reducing and calibrating the OAJ data used in this work.
Funding for OAJ, UPAD, and CEFCA has been provided by the Governments of 
Spain and Arag\'on through the Fondo de Inversiones de Teruel; the 
Aragonese Government through the Research Groups E96, E103, E16\_17R, 
and E16\_20R; the Spanish Ministry of Science, Innovation and 
Universities (MCIU/AEI/FEDER, UE) with grant PGC2018-097585-B-C21; the 
Spanish Ministry of Economy and Competitiveness (MINECO/FEDER, UE) under 
AYA2015-66211-C2-1-P, AYA2015-66211-C2-2, AYA2012-30789, and 
ICTS-2009-14; and European FEDER funding (FCDD10-4E-867, 
FCDD13-4E-2685).
Funding for the J-PAS Project has been provided by the Governments of Spain and Aragón through the Fondo de Inversi{\'o}n de Teruel, European FEDER funding and the Spanish Ministry of Science, Innovation and Universities, and by the Brazilian agencies FINEP, FAPESP, FAPERJ and by the National Observatory of Brazil. Additional funding was also provided by the Tartu Observatory and by the J-PAS Chinese Astronomical Consortium.
This work has made use of data from the European Space Agency (ESA) mission {\it Gaia} (\url{https://www.cosmos.esa.int/gaia}), processed by the {\it Gaia} Data Processing and Analysis Consortium (DPAC, \url{https:// www.cosmos.esa.int/web/gaia/dpac/ consortium}). Funding for the DPAC has been provided by national institutions, in particular the institutions participating in the {\it Gaia} Multilateral Agreement. Guoshoujing Telescope (the Large Sky Area Multi-Object Fiber Spectroscopic Telescope LAMOST) is a National Major Scientific Project built by the Chinese Academy of Sciences. Funding for the project has been provided by the National Development and Reform Commission. LAMOST is operated and managed by the National Astronomical Observatories, Chinese Academy of Sciences.

\section*{Data Availability}
The data underlying this article is available online via \url{http://www.j-pas.org/datareleases/minijpas_public_data_release_pdr201912} and 
\url{http://www.j-plus.es/ancillarydata/index}.
The VOSA temperatures as well as the metallicities in Section 5 are available from the authors upon request.



%\begin{thebibliography}{99}
%\end{thebibliography}

%%%%%%%%%%%%%%%%%%%%%%%%%%%%%%%%%%%%%%%%%%%%%%%%%%

%%%%%%%%%%%%%%%%% APPENDICES %%%%%%%%%%%%%%%%%%%%%

%%%%%%%%%%%%%%%%%%%%%%%%%%%%%%%%%%%%%%%%%%%%%%%%%%

% Don't change these lines
\bsp	% typesetting comment
\label{lastpage}
\end{document}